\documentclass[12pt,eqno,epsf]{article}
\usepackage{amsmath,amssymb,graphicx,slashbox}
\numberwithin{equation}{section}

\def\mydate{July 1, 2014}
\def\ignore#1{{}}

\newcounter{sxn}

\newcounter{axn}

\date{}

\newdimen\mybaselineskip
\renewcommand{\thefootnote}{\arabic{footnote}}

\newcommand{\beeq}{\begin{equation}}
\newcommand{\eneq}{\end{equation}}
\newcommand{\beqn}{\begin{eqnarray}}
\newcommand{\eeqn}{\end{eqnarray}}

\newcommand{\alp}{\alpha}
\newcommand{\bt}{\beta}
\newcommand{\gm}{\gamma}
\newcommand{\Gm}{\Gamma}
\newcommand{\dlt}{\delta}

\newcommand{\ep}{\epsilon}

\newcommand{\tht}{\theta}
\newcommand{\thH}{\theta_{\rm H}}

\newcommand{\lmd}{\lambda}

\newcommand{\sgm}{\sigma}

\newcommand{\vph}{\varphi}
\newcommand{\omg}{\omega}

\newcommand{\be}{\begin{equation}}
\newcommand{\ee}{\end{equation}}
\newcommand{\bea}{\begin{eqnarray}}
\newcommand{\eea}{\end{eqnarray}}
\newcommand{\eql}{\!\!\!&=\!\!\!&}

\newcommand{\defa}{\!\!\!&\equiv\!\!\!&}

\newcommand{\simlt}{\stackrel{<}{{}_\sim}}

\newcommand{\tl}[1]{\tilde{#1}}
\newcommand{\bdm}[1]{{\mbox{\boldmath $#1$}}}
\newcommand{\tr}{{\rm tr}}

\newcommand{\diag}{{\rm diag}}
\newcommand{\der}{\partial}
\newcommand{\dr}{\!\!d}
\newcommand{\hc}{{\rm h.c.}}
\newcommand{\ie}{{i.e.}}
\newcommand{\id}{\mbox{\boldmath $1$}}
\newcommand{\sgn}{{\rm sgn}}
\newcommand{\zf}{z_{\rm f}}

\newcommand{\vev}[1]{\langle #1 \rangle}

\newcommand{\brkt}[1]{\left( #1 \right)}
\newcommand{\brc}[1]{\left\{ #1 \right\}}
\newcommand{\sbk}[1]{\left[ #1 \right]}
\newcommand{\abs}[1]{\left| #1 \right|}

\renewcommand{\Im}{{\rm Im}\,}

\newcommand{\cA}{{\cal A}}
\newcommand{\cB}{{\cal B}}

\newcommand{\cD}{{\cal D}}

\newcommand{\cH}{{\cal H}}

\newcommand{\cL}{{\cal L}}

\newcommand{\cO}{{\cal O}}
\newcommand{\cP}{{\cal P}}
\newcommand{\cQ}{{\cal Q}}
\newcommand{\cR}{{\cal R}}

\begin{document}
\thispagestyle{empty}

\baselineskip=12pt

{\small \noindent \mydate    
\hfill }

{\small \noindent \hfill  KEK-TH-1748}

\baselineskip=35pt plus 1pt minus 1pt

\vskip 1.5cm

\begin{center}
{\Large \bf 6D gauge-Higgs unification on $T^2/Z_N$}\\
{\Large \bf with custodial symmetry}\\

\vspace{1.5cm}
\baselineskip=20pt plus 1pt minus 1pt

\normalsize

{\bf Yoshio\ Matsumoto},${}^1\!${\def\thefootnote{\fnsymbol{footnote}}
\footnote[1]{\tt e-mail address: yoshio@post.kek.jp}} 
{\bf and 
Yutaka\ Sakamura}${}^{1,2}\!${\def\thefootnote{\fnsymbol{footnote}}
\footnote[2]{\tt e-mail address: sakamura@post.kek.jp}}

\vspace{.3cm}
${}^1${\small \it Department of Particles and Nuclear Physics, \\
The Graduate University for Advanced Studies (Sokendai), \\ 
Tsukuba, Ibaraki 305-0801, Japan} \\ \vspace{3mm}
${}^2${\small \it KEK Theory Center, Institute of Particle and Nuclear Studies, 
KEK, \\ Tsukuba, Ibaraki 305-0801, Japan} 
\end{center}

\vskip 1.0cm
\baselineskip=20pt plus 1pt minus 1pt

\begin{abstract}
We investigate the gauge-Higgs unification models compactified on $T^2/Z_N$ 
that have the custodial symmetry. 
We select possible gauge groups, orbifolds and representations 
of the matter fermions that are consistent with the custodial symmetry, 
by means of the group theoretical analysis. 
The best candidate we found is 6D ${\rm SU(3)}_C\times{\rm U(4)}$ gauge theory 
on $T^2/Z_3$ and the third generation quarks are embedded into bulk fermions 
that are the symmetric traceless rank-2 tensor of SO(6). 
\end{abstract}


\newpage

\section{Introduction}
The gauge-Higgs unification~\cite{Manton:1979kb,Fairlie:1979at,Hosotani:1983xw,
Hatanaka:1998yp} 
is an interesting candidate for the new physics 
beyond the standard model. 
This solves the gauge hierarchy problem because 
a higher dimensional gauge symmetry protects the Higgs mass 
against quantum corrections. 
Chiral fermions in four-dimensional (4D) effective theory can be obtained 
by compactifying the extra dimensions on an orbifold. 

The simplest models of this category are based on 
five-dimensional (5D) gauge theories whose gauge groups are 
U(3) in the flat spacetime~\cite{Scrucca:2003ut,Cacciapaglia:2005da,Panico:2005dh}, 
and ${\rm SO(5)}\times{\rm U(1)}$ 
in the warped spacetime~\cite{Agashe:2004rs,Hosotani:2006qp,Hosotani:2008tx}. 
In these models, the electroweak symmetry is broken by the vacuum expectation value (VEV) 
of the Wilson line phase~$\thH\equiv\int_C dy\,A_y$, 
where $C$ is a non-contractible cycle along the extra dimension 
and $A_y$ is the extra-dimensional component of the gauge field. 
The mass scale of the Kaluza-Klein (KK) excitation modes~$m_{\rm KK}$ 
is determined by $\vev{\thH}$ as 
$m_{\rm KK}\simeq m_W/\abs{\vev{\thH}}$ ($m_W$: the W boson mass) in the flat spacetime,  
and $m_{\rm KK}\simeq m_W\pi\sqrt{k\pi R}/\abs{\sin\vev{\thH}}$ 
($e^{k\pi R}$: the warp factor) in the warped spacetime~\cite{Hosotani:2006qp}. 
The current experimental constraints require $\vev{\thH}$ to be small, 
\ie, $\vev{\thH}\simlt\cO(0.1)$. 
In order to realize such small values of $\vev{\thH}$, 
we need some amount of fine-tuning among the model parameters, 
such as 5D mass parameters for fermions. 
This stems from the fact that the effective potential for $\thH$ 
does not exist at tree level and is induced at one-loop level 
in 5D gauge-Higgs unification models.  
The one-loop potential is typically expressed 
as a sum of periodic functions of $\thH$ 
with the period of $\pi$ and $2\pi$ (or $\pi/2$), 
which are roughly approximated as 
the cosine functions~\cite{Contino:2006qr,Sakamura:2010ju}. 
Therefore, $\vev{\thH}=\cO(1)$ is realized 
unless the model parameters are fine-tuned. 
This problem can be evaded in the six-dimensional (6D) gauge-Higgs unification models. 
In this case, quartic terms in the Wilson line phases exist at tree level, 
while quadratic terms are induced at one-loop level. 
In the flat spacetime, for example, 
the effective potential has a form of 
\be
 V(\thH) = -\frac{c_2g^2}{l_6R^2}\brkt{\frac{\thH}{g\pi R}}^2
 +c_4g^2\brkt{\frac{\thH}{g\pi R}}^4+\cO(\thH^6), 
\ee
where $c_2,c_4=\cO(1)$ are numerical constants, 
$g$ is the ${\rm SU(2)}_{\rm L}$ gauge coupling constant, 
$l_6\equiv 128\pi^3$ is the 6D loop factor, 
and $R$ is a typical radius of the extra-dimensional space. 
By minimizing this, we find that 
\be
 \vev{\thH} \simeq \frac{g\pi\sqrt{c_2}}{\sqrt{2l_6c_4}} 
 \simeq \frac{0.02\sqrt{c_2}}{\sqrt{c_4}} \ll 1, 
\ee
and the KK modes are estimated to be around a few TeV.  

In extra-dimensional models, coupling constants in 4D 
effective theories generally deviate from the standard model values 
even at tree level due to mixing 
with the KK modes~\cite{Sakamura:2006rf,Hosotani:2007qw,Sakamura:2007qz}. 
Unless $m_{\rm KK}$ is very high, models need some mechanisms 
to suppress such deviations. 
Especially a requirement that 
the $\rho$ parameter and the Z boson coupling to the left-handed bottom quark 
(the $Zb_L\bar{b}_L$ coupling) do not deviate too much 
often imposes severe constraints on the model building. 
It is known that the custodial symmetry 
can protect them against the corrections induced by the mixing 
with the KK modes~\cite{Agashe:2004rs,Agashe:2006at}. 
Hence we focus on 6D gauge-Higgs unification models that has 
the custodial symmetry in this paper. 

The purpose of this paper is to select candidates for realistic 
6D gauge-Higgs unification 
models by means of the group theoretical analysis. 
The analysis is useful to investigate 
the gauge-Higgs unification models because the Higgs sector is determined 
by the gauge group structure. 
There are some works along this direction. 
5D models are analyzed in Ref.~\cite{Grzadkowski:2006tp}, 
the tree-level Higgs potentials in 6D models are calculated 
in Ref.~\cite{Chang:2012iq}, 
and models in arbitrary dimensions 
are discussed in Ref.~\cite{Aranda:2010hn}. 
In these works, the custodial symmetry is not considered 
and the electroweak gauge symmetry~${\rm SU(2)}_{\rm L}\times {\rm U(1)}_Y$ 
is embedded into a simple group. 
Thus the Weinberg angle~$\tht_W$ is determined only by the group structure, 
and they found that no simple group realizes the observed value of $\tht_W$. 
However, the assumption that ${\rm SU(2)}_{\rm L}\times {\rm U(1)}_Y$ 
is embedded into a simple group is not indispensable  
because the color symmetry~SU(3)${}_C$ is not unified anyway. 
Besides, any brane localized terms allowed by the symmetries 
are not introduced in Refs.~\cite{Grzadkowski:2006tp,Aranda:2010hn}. 
In fact, the realistic models constructed so far allow 
both an extra U(1) gauge symmetry, 
which is relevant to the realization of the experimental value of $\tht_W$, 
and various terms and fields localized at the fixed points of the 
orbifolds~\cite{Cacciapaglia:2005da,Panico:2005dh,Hosotani:2008tx,Contino:2006qr}. 
Therefore, we include both ingredients in our analysis. 
Since larger gauge groups contain more unwanted exotic particles, 
we consider a case that the 6D gauge group is ${\rm SU(3)}_C\times G\times{\rm U(1)}$, 
where $G$ is a simple group whose rank is less than four. 

The paper is organized as follows. 
In the next section, we explain our setup and derive conditions for zero-modes. 
In Sec.~\ref{ZM:gauge}, we list the zero-modes in the bosonic sector 
for all the rank-two and the rank-three groups that include the custodial symmetry. 
In Sec.~\ref{CW}, we find a condition to preserve the custodial symmetry, 
and provide explicit expressions of the W and Z boson masses. 
In Sec.~\ref{bulk_fermion}, we discuss embeddings of quarks into 6D fermions, 
and search for appropriate representations of $G$ that the 6D fermions should belong to. 
In Sec.~\ref{HiggsPotential}, we calculate the Higgs potential at tree level. 
Sec.~\ref{summary} is devoted to the summary. 
In Appendix~\ref{CartanWeyl}, 
we collect formulae in the Cartan-Weyl basis of the gauge group generators. 
In Appendix~\ref{general_obBC}, 
general forms of the orbifold boundary conditions are shown. 
In Appendix~\ref{decompList}, we list irreducible decompositions 
of various $G$ representations 
into the ${\rm SU(2)}_{\rm L}\times{\rm SU(2)}_{\rm R}$ multiplets.

\section{Setup}
\subsection{Compactified space}
The 6D spacetime is assumed to be flat, and the metric is given by 
\be
 ds^2 = \eta_{MN}dx^M dx^N = \eta_{\mu\nu}dx^\mu dx^\nu+(dx^4)^2+(dx^5)^2, 
\ee
where $M,N=0,1,\cdots,5$, 
$\eta_{\mu\nu}=\diag(-1,1,1,1)$ is the 4D Minkowski metric, 
and a point in the extra space~$(x^4,x^5)$ is identified as 
\be
 \begin{pmatrix} x^4 \\ x^5 \end{pmatrix} 
 \sim \begin{pmatrix} x^4 \\ x^5 \end{pmatrix}
 +2\pi n_1R_1\begin{pmatrix} 1 \\ 0 \end{pmatrix}
 +2\pi n_2R_2\begin{pmatrix} \cos\tht \\ \sin\tht \end{pmatrix}, 
\ee
where $n_1$ and $n_2$ are integers, 
and $R_1,R_2>0$ and $0<\tht<\pi$ are constants. 
In order to obtain a 4D chiral theory at low energies, 
we compactify the extra space on a two-dimensional orbifold. 
All possible orbifolds are $T^2/Z_N$ ($N=2,3,4,6$)~\cite{LecNotes}. 
It is convenient to use a complex (dimensionless) 
coordinate~$z\equiv\frac{1}{2\pi R_1}(x^4+ix^5)$. 
Then, the orbifold obeys the identification, 
\be
 z \sim \omg z+n_1+n_2\tau, 
\ee
where $\omg=e^{2\pi i/N}$ and $\tau\equiv\frac{R_2}{R_1}e^{i\tht}$. 
Note that an arbitrary value of $\tau$ is allowed when $N=2$ 
while it must be equal to $\omg$ when $N\neq 2$. 

The orbifold~$T^2/Z_N$ has the following fixed points 
in the fundamental domain~\cite{Kobayashi:1991rp,Abe:2013bca}. 
\be
 z = \zf \equiv \begin{cases} 0, \frac{1}{2},\frac{\tau}{2},\frac{1+\tau}{2} & 
 (\mbox{on $T^2/Z_2$}) \\
 0, \frac{2+\tau}{3}, \frac{1+2\tau}{3} &
 (\mbox{on $T^2/Z_3$}) \\
 0, \frac{1+\tau}{2} & 
 (\mbox{on $T^2/Z_4$}) \\
 0 & (\mbox{on $T^2/Z_6$}) \end{cases}
\ee
4D fields or interactions are allowed to be introduced
on these fixed points.

\subsection{Field content}
We consider a 6D gauge theory whose gauge group is 
${\rm SU(3)}_C\times G\times {\rm U(1)}_Z$, 
where $G$ is a simple group. 
Since $G$ must include ${\rm SU(2)}_{\rm L}\times {\rm SU(2)}_{\rm R}$, 
its rank~$r$ is greater than one. 
In this paper, we investigate cases of $r=2,3$. 
In the following, we omit ${\rm SU(3)}_C$ since it is irrelevant to the discussion. 
The 6D gauge fields for $G$ and ${\rm U(1)}_Z$ are denoted as $A_M$ and $B_M^Z$, 
and the field strengths and the covariant derivative 
are defined as $F^{(A)}_{MN}\equiv\der_M A_N-\der_N A_M-i[A_M,A_N]$, 
$F^{(Z)}_{MN}\equiv\der_M B^Z_N-\der_N B^Z_M$, 
and $\cD_M\equiv\der_M-iA_M-iq_ZB^Z_M$, 
where $q_Z$ is a ${\rm U(1)}_Z$ charge.  
The 6D Lagrangian is expressed as 
\bea
 \cL \eql -\frac{1}{4g_A^2}\tr\brkt{F^{(A)MN}F^{(A)}_{MN}}
 -\frac{1}{4g_Z^2}F^{(Z)MN}F^{(Z)}_{MN}
 +i\sum_f\bar{\Psi}^f\Gm^M\cD_M\Psi^f  \nonumber\\
 &&+\sum_{\zf}\cL^{(\zf)}\dlt^{(2)}(z-\zf), 
\eea
where $g_A$ and $g_Z$ are the 6D gauge coupling constants 
for $G$ and ${\rm U(1)}_Z$, $\Gm^M$ are the 6D gamma matrices, 
and $\cL^{(\zf)}$ are 4D Lagrangians localized at the fixed points~$z=\zf$. 

The $G$ gauge field~$A_M$ is decomposed as
\be
 A_M = \sum_i C^i_M H_i+\sum_\alp W^\alp_M E_\alp, \label{decomp:A_M}
\ee
where $\{H_i,E_\alp\}$ are the generators in the Cartan-Weyl basis, \ie, 
$H_i$ ($i=1,\cdots,r$) are the Cartan generators and $\alp$ runs over 
all the roots of $G$. 
Since $A_M$ is Hermitian, $C_M^i$ are real and $W_M^{-\alp}=(W_M^\alp)^*$. 
In the complex coordinate~$(x^\mu,z)$, 
the extra-dimensional components of the gauge fields are expressed as 
\bea
 A_z \eql \pi R_1\brkt{A_4-iA_5}, \;\;\;\;\;
 A_{\bar{z}} = A_z^\dagger, \nonumber\\
 B_z^Z \eql \pi R_1\brkt{B_4^Z-iB_5^Z}, \;\;\;\;\;
 B_{\bar{z}}^Z = B_z^{Z\dagger}. 
\eea

\ignore{
We assume a nonvanishing ${\rm U(1)}_Z$ magnetic flux on $T^2/Z_N$ as a background 
while no ${\rm U(1)}_F$ flux. 
\be
 \cB \equiv \int_{T^2/Z_N}dzd\bar{z}\;\vev{F_{z\bar{z}}^Z} 
 = -\frac{2i\Im\tau}{N}\vev{F_{z\bar{z}}^Z} 
 = \cA\vev{F_{45}^Z},  
\ee
where we have used~\footnote{
The $z\bar{z}$-component of the field strength is defined as
$F_{z\bar{z}}^{(Z)} \equiv \der_z B^Z_{\bar{z}}-\der_{\bar{z}}B^Z_z 
= 2i(\pi R_1)^2F_{45}^{(Z)}$. } 
that $\vev{F_{z\bar{z}}}$ is a constant at the second equality, 
and $\cA\equiv (2\pi R_1)^2\Im\tau/N=4\pi^2R_1R_2\sin\tht/N$ 
is the area of the fundamental domain of $T^2/Z_N$. 
Then the vector potential~$B_z^Z$ can have a nontrivial background,
\be
 \vev{B_z^Z} = -\frac{iN\cB(\bar{z}+\bar{a}_w)}{4\Im\tau},  
 \label{bg:B_z^Z}
\ee
where a complex constant~$a_w$ is a complex Wilson line phase, 
which can be absorbed into the Scherk-Schwarz phases 
(see (\ref{def:newSSphases})). 
Note that the Wilson line phase~$a_w$ is not a dynamical field 
because $B_z^Z$ does not have a zero-mode 
as we will see in the next subsection. 
The magnetic flux~$\cB$ is quantized 
as we will explain in Sec.~\ref{bulk_fermion} 
(see (\ref{cB:quantize})). 
}

\subsection{Orbifold conditions for gauge fields}
As shown in Appendix~\ref{general_obBC}, 
the general orbifold boundary conditions for the gauge fields can be expressed as  
\bea
 A_M(x,z+1) \eql A_M(x,z), \;\;\;\;\;
 B_M^Z(x,z+1) = B_M^Z(x,z), \nonumber\\
 A_M(x,z+\tau) \eql A_M(x,z), \;\;\;\;\;
 B_M^Z(x,z+\tau) = B_M^Z(x,z), \nonumber\\
 A_\mu(x,\omg z) \eql P A_\mu(x,z) P^{-1}, \;\;\;\;\;
 A_z(x,\omg z) = \omg^{-1}P A_z(x,z)P^{-1}, \nonumber\\
 B_\mu^Z(x,\omg z) \eql B_\mu^Z(x,z), \;\;\;\;\;
 B_z^Z(x,\omg z) = \omg^{-1}B_z^Z(x,z), \label{orbifold_BC:gauge}
\eea
where $P$ is an element of $G$. 
The orbifold conditions for 6D fermions are provided in (\ref{ob_cond:fermion}). 

Since zero-modes of the gauge fields have flat profiles 
over the extra dimensional space, 
we can see from (\ref{orbifold_BC:gauge}) that 
$B_\mu^Z$ has a zero-mode while $B_z^Z$ does not. 
Namely ${\rm U(1)}_Z$ is unbroken by the orbifold conditions. 
The condition for $A_M$ to have zero-modes is determined 
by the choice of the matrix~$P$ in (\ref{orbifold_BC:gauge}). 
It is always possible to choose the generators 
so that $P$ is expressed as
\be
 P = \exp\brkt{ip\cdot H}, 
\ee
where $p\cdot H\equiv \sum_ip_iH_i$ and $p_i$ are real constants. 
Thus $PH_iP^{-1}=H_i$ and $PE_\alp P^{-1}=e^{ip\cdot\alp}E_\alp$, 
and the relevant conditions in (\ref{orbifold_BC:gauge}) 
to the zero-mode conditions are rewritten as  
\bea
 C_\mu^i(x,\omg z) \eql C_\mu^i(x,z), \;\;\;\;\;
 C_z^i(x,\omg z) = \omg^{-1}C_z^i(x,z), \nonumber\\
 W_\mu^\alp(x,\omg z) \eql e^{ip\cdot\alp}W_\mu^\alp(x,z), \;\;\;\;\;
 W_z^\alp(x,\omg z) = e^{i(p\cdot\alp-\frac{2\pi}{N})}W_z^\alp(x,z). 
 \label{ob_BC:gauge2}
\eea
This indicates that $C_\mu^i$ always have zero-modes 
while $C_z^i$ do not irrespective of the choice of the matrix~$P$. 
Therefore the orbifold boundary conditions cannot reduce the rank of $G$ 
as pointed out in Ref.~\cite{Hebecker:2001jb}. 
In contrast, whether $W_\mu^\alp$ and $W_z^\alp$ have zero-modes depend 
on the choice of $P$. 
Since (\ref{ob_BC:gauge2}) is the $Z_N$ transformation, $p_i$ must satisfy 
$e^{iNp\cdot\alp}=\id$. 
Thus possible values of $p\cdot\alp$ are 
\be
 p\cdot\alp = \frac{2n_\alp\pi}{N}, \label{value:palp}
\ee
where $n_\alp$ is an integer. 
\ignore{
From (\ref{ob_BC:gauge2}), 
the conditions for $W_\mu^\alp$ and $W_z^\alp$ to have zero-modes 
are expressed as 
\bea
 p\cdot\alp = \begin{cases} 0, & (\mbox{for $W_\mu^\alp$}) \\
 \frac{2\pi}{N}, & (\mbox{for $W_z^\alp$}) \end{cases}
\eea
where the equalities hold modulo $2\pi$. 
}

In this paper, we focus on $P$ such that the orbifold bondary conditions break $G$ 
to ${\rm SU(2)}_{\rm L}\times {\rm SU(2)}_{\rm R}\times {\rm U(1)}^{r-2}$. 
We denote the positive roots that specify ${\rm SU(2)}_{\rm L}$ 
and ${\rm SU(2)}_{\rm R}$ as 
$\alp_L$ and $\alp_R$, respectively. 
The ${\rm SU(2)}_{\rm L}$ and ${\rm SU(2)}_{\rm R}$ generators are 
given by (\ref{def:T_LR}). 
Then (\ref{value:palp}) is further restricted as  
\bea
 p\cdot\alp_L \eql p\cdot\alp_R = 0, \;\;\;\;\; (\mbox{mod $2\pi$}) \nonumber\\
 p\cdot\bt \eql \frac{2n_\bt\pi}{N}. \;\;\;\;\; 
 (\bt\neq\alp_L,\alp_R, \;\;\;
 n_\bt\in\mathbb{Z}, \;\;\; n_\bt\not\in N\mathbb{Z}) 
 \label{zm_cond:gauge}
\eea
From the last condition in (\ref{ob_BC:gauge2}),
the zero-mode condition for $W_z^\bt$ is 
\be
 p\cdot\bt = \frac{2\pi}{N}. \label{zm_cond:scalar}
\ee

\section{Zero-modes of gauge and Higgs fields} \label{ZM:gauge}
In this section, we investigate the field content 
of the zero-modes from the 6D gauge fields. 

\subsection{Rank-two groups}
First we consider a case of $r=2$, \ie, $G={\rm SO(5)},{\rm G}_2$. 
In this case, the unbroken gauge group by the orbifold conditions 
is ${\rm SU(2)}_{\rm L}\times {\rm SU(2)}_{\rm R}\times {\rm U(1)}_Z$. 
We do not consider $G=$SU(3) because it does not contain 
${\rm SU(2)}_{\rm L}\times {\rm SU(2)}_{\rm R}$ as a subgroup. 
The roots of $G$ can be expressed as linear combinations of 
two-dimensional basis vectors~$\bdm{e}^i$ ($i=1,2$). 

\subsubsection{SO(5)} \label{zm:SO5}
The roots are $\{\pm\bdm{e}^i\pm\bdm{e}^j,\pm\bdm{e}^i\}$ ($1\leq i\neq j\leq 2$). 
We can choose the unbroken subgroup~${\rm SU(2)}_{\rm L}\times {\rm SU(2)}_{\rm R}$ as 
\be
 (\alp_L,\alp_R) = (\bdm{e}^1+\bdm{e}^2,\bdm{e}^1-\bdm{e}^2). 
\ee
The other possible choices are essentially equivalent to this case.\footnote{
We cannot choose them as $(\alp_L,\alp_R)=(\bdm{e}^1,\bdm{e}^2)$ 
because $\alp_L+\alp_R$ is a root in such a case. 
}
Then the adjoint representation of $G$ is decomposed into 
the irreducible representations of ${\rm SU(2)}_{\rm L}\times {\rm SU(2)}_{\rm R}$ as
\be
 \bdm{10 = (3,1)+(1,3)+(2,2)}.  \label{IRdecomp:SO5}
\ee
A candidate for the Higgs fields is a bidoublet~$\bdm{(2,2)}$, 
which consists of $\pm\bdm{e}^1$ and $\pm\bdm{e}^2$. 
The conditions in (\ref{zm_cond:gauge}) are now expressed as 
\bea
 p_1+p_2 \eql p_1-p_2 = 0, \;\;\;\;\; (\mbox{mod $2\pi$}) \nonumber\\
 p_1 \eql \frac{2n_P\pi}{N}. \;\;\;\;\; (n_P\in\mathbb{Z}, \;\;\; n_P\not\in N\mathbb{Z}) 
\eea
It is enough to find a solution in a range:~$0\leq p_1,p_2<2\pi$. 
A solution exists when $N\neq 3$, and it is
\be
 (p_1,p_2) = (\pi,\pi), 
\ee
or
\be
 P = \exp\brc{i\pi(H_1+H_2)}. 
\ee
Therefore the zero-mode condition~(\ref{zm_cond:scalar}) for $\bdm{(2,2)}$ 
is expressed as 
\be
 \pi = \frac{2\pi}{N}. 
\ee
Namely, we have one Higgs bidoublet when $N=2$, 
while no Higgs exists in the other cases.

\subsubsection{G$\bdm{{}_2}$}
The roots are $\{\pm(\bdm{e}^1\pm\sqrt{3}\bdm{e}^2)/2, 
\pm(\bdm{e}^1\pm\frac{1}{\sqrt{3}}\bdm{e}^2)/2,\pm\bdm{e}^1,  
\pm\bdm{e}^2/\sqrt{3}\}$. 
We can choose the ${\rm SU(2)}_{\rm L}\times {\rm SU(2)}_{\rm R}$ subgroup as
\bea
 (\alp_L,\alp_R) \eql \brkt{\bdm{e}^1,\frac{\bdm{e}^2}{\sqrt{3}}}, \;\;\;
 \brkt{\frac{\bdm{e}^2}{\sqrt{3}},\bdm{e}^1}. 
\eea
The other possible choices are essentially equivalent to these cases. 

Let us first consider the case of $(\alp_L,\alp_R)=(\bdm{e}^1,\bdm{e}^2/\sqrt{3})$. 
The irreducible decomposition of the adjoint representation of $G$ is
\be
 \bdm{14 = (3,1)+(1,3)+(2,4)}. \label{IRdecomp:G2}
\ee
A candidate for the Higgs fields is $\bdm{(2,4)}$. 
The conditions in (\ref{zm_cond:gauge}) become 
\bea
 &&p_1 = \frac{p_2}{\sqrt{3}} = 0, \;\;\;\;\; (\mbox{mod $2\pi$}) \nonumber\\
 &&\frac{p_1}{2}+\frac{p_2}{2\sqrt{3}} = \frac{2n_P\pi}{N}. \;\;\;\;\;
 (n_P\in\mathbb{Z}, \;\;\; n_P\not\in N\mathbb{Z})
\eea
It is enough to find a solution in a range~$0\leq p_1,\frac{p_2}{\sqrt{3}}<2\pi$. 
A solution exists when $N\neq 3$, and it is 
\be
 P = \exp\brkt{2\sqrt{3}\pi iH_2}. 
\ee
Therefore the zero-mode condition~(\ref{zm_cond:scalar}) for $\bdm{(2,4)}$ 
is expressed as 
\be
 \pi = \frac{2\pi}{N}. 
\ee
Namely, we have a $\bdm{(2,4)}$ multiplet as the Higgs fields 
when $N=2$, while no Higgs exists in the other cases. 

In the case of $(\alp_L,\alp_R)=(\bdm{e}^2/\sqrt{3},\bdm{e}^1)$,  
the results are obtained by exchanging ${\rm SU(2)}_{\rm L}$ 
and ${\rm SU(2)}_{\rm R}$ in the above resuts. 
Hence we do not have ${\rm SU(2)}_{\rm L}$-doublet Higgses.

\subsection{Rank-three groups} \label{ZM:rank3}
Next we consider a case of $r=3$, \ie, $G=$SU(4),SO(7),Sp(6). 
In this case, the unbroken gauge group by the orbifold conditions is 
${\rm SU(2)}_{\rm L}\times {\rm SU(2)}_{\rm R}\times {\rm U(1)}_X\times {\rm U(1)}_Z$. 
The roots of $G$ can be expressed as linear combinations of 
three-dimensional basis vectors~$\bdm{e}^i$ ($i=1,2,3$). 

\subsubsection{SU(4)} \label{zm:SU4}
The roots are $\{\sqrt{2}\bdm{e}^1$, $\sqrt{2}\bdm{e}^2$, 
$\pm\frac{\bdm{e}^1}{\sqrt{2}}\pm\frac{\bdm{e}^2}{\sqrt{2}}+\bdm{e}^3\}$.\footnote{
It is sometimes convenient to embed these roots 
into a four-dimensional vector space. 
Then they are expressed as $\bdm{\hat{e}}^I-\bdm{\hat{e}}^J$ 
($1\leq I\neq J\leq 4$), where $\bdm{\hat{e}}^I$ are the basis vectors 
of the embeded space. 
The original basis vectors are expressed as
$\bdm{e}^1=\frac{1}{\sqrt{2}}(\bdm{\hat{e}}^1-\bdm{\hat{e}}^2)$, 
$\bdm{e}^2=\frac{1}{\sqrt{2}}(\bdm{\hat{e}}^3-\bdm{\hat{e}}^4)$ and 
$\bdm{e}^3=\frac{1}{2}(\bdm{\hat{e}}^1+\bdm{\hat{e}}^2-\bdm{\hat{e}}^3-\bdm{\hat{e}}^4)$. 
} 
We can choose the ${\rm SU(2)}_{\rm L}\times {\rm SU(2)}_{\rm R}$ subgroup as
\be
 (\alp_L,\alp_R) = (\sqrt{2}\bdm{e}^1,\sqrt{2}\bdm{e}^2). 
\ee
The other choices are essentially equivalent to this case. 
The $U(1)_X$ generator~$Q_X$ is identified as 
\be
 Q_X = 2\bdm{e}_3\cdot H = 2H_3. 
\ee
The irreducible decomposition of the adjoint representation of $G$ is
\be
 \bdm{15 = (3,1)_0+(1,3)_0+(2,2)_{+2}+(2,2)_{-2}+(1,1)_0}, \label{adj_decomp:SU4}
\ee
where $\bdm{(3,1)_0}$, $\bdm{(1,3)_0}$ and $\bdm{(1,1)_0}$ correspond to 
${\rm SU(2)}_{\rm L}$, ${\rm SU(2)}_{\rm R}$ and ${\rm U(1)}_X$ generators, respectively. 
Thus the candidates for the Higgs fields are two bidoublets. 
The conditions in (\ref{zm_cond:gauge}) become 
\bea
 &&\sqrt{2}p_1 = \sqrt{2}p_2 = 0, \;\;\;\;\; (\mbox{mod $2\pi$}) \nonumber\\
 &&\frac{p_1}{\sqrt{2}}+\frac{p_2}{\sqrt{2}}+p_3 = \frac{2n_P\pi}{N}, 
 \;\;\;\;\; (n_P\in\mathbb{Z}, \;\; n_P\not\in N\mathbb{Z})
\eea
Solutions are  
\be
 P = \exp\brkt{\frac{2n_P\pi i}{N}H_3},  
\ee
where $n_P=1,\cdots,N-1$. 
Therefore the zero-mode conditions~(\ref{zm_cond:scalar}) for $\bdm{(2,2)_{\pm 2}}$ are
\be
 \pm\frac{2n_P\pi}{N} = \frac{2\pi}{N}. \;\;\;\;\; (\mbox{mod $2\pi$}) 
\ee
Namely, the scalar zero-modes we have are 
\bea
 \bdm{(2,2)_{+2}}, \;\; \bdm{(2,2)_{-2}} &:& (\mbox{when $N=2$}) \nonumber\\
 \bdm{(2,2)_{+2}} &:& (\mbox{when $N=3,4,6$ and $n_P=1$}) \nonumber\\
 \bdm{(2,2)_{-2}} &:& (\mbox{when $N=3,4,6$ and $n_P=N-1$}) \nonumber\\
 \mbox{Nothing} &:& (\mbox{in the other cases}) 
\eea

\ignore{
Although the roots are three-dimensional vectors in this case, it is convenient 
to embed them into the four-dimensional vector space 
whose basis vectors are $\bdm{\hat{e}}^I$ ($I=1,2,3,4$). 
The root space is specified as the orthogonal space to 
$\bdm{f}\equiv\bdm{\hat{e}}^1+\bdm{\hat{e}}^2+\bdm{\hat{e}}^3+\bdm{\hat{e}}^4$. 
The relation between the bases~$\{\bdm{e}^i\}$ and $\{\bdm{\hat{e}}^I\}$ is given 
by (\ref{embed_basis}) and (\ref{4to3projection}). 
The embeded 4-component vector~$\hat{p}_I$ for $p_i$ is defined as
$(\hat{p}_1,\hat{p}_2,\hat{p}_3,\hat{p}_4)\equiv (p_1,p_2,p_3)\cP^t$. 
The roots are then expressed as $\bdm{\hat{e}}^I-\bdm{\hat{e}}^J$ 
($1\leq I\neq J\leq 4$). 
We can choose the ${\rm SU(2)}_{\rm L}\times {\rm SU(2)}_{\rm R}$ subgroup as
\be
 (\alp_L,\alp_R) = (\bdm{\hat{e}}^1-\bdm{\hat{e}}^2,\bdm{\hat{e}}^3-\bdm{\hat{e}}^4), 
\ee
The other possible choices are essentially equivalent to this case. 
The unbroken $U(1)_X$ is specified by a vector orthogonal to 
$\alp_L$, $\alp_R$ and $\bdm{f}$, \ie, 
$\bdm{\hat{e}}^1+\bdm{\hat{e}}^2-\bdm{\hat{e}}^3-\bdm{\hat{e}}^4$.  
Namely the $U(1)_X$ generator is 
\be
 Q_X = \brkt{1,1,-1,-1}\cP\cdot H, 
\ee
where $\cP$ is the projection matrix given by (\ref{4to3projection}). 
The irreducible decomposition of the adjoint representation of $G$ is
\be
 \bdm{15 = (3,1)_0+(1,3)_0+(2,2)_{+2}+(2,2)_{-2}+(1,1)_0}. 
\ee
where $\bdm{(3,1)_0}$, $\bdm{(1,3)_0}$ and $\bdm{(1,1)_0}$ correspond to 
${\rm SU(2)}_{\rm L}$, ${\rm SU(2)}_{\rm R}$ and ${\rm U(1)}_X$ generators, respectively. 
Thus the candidates for the Higgs scalars are two bidoublets. 
Independent conditions in (\ref{zm_cond:gauge}) are expressed as
\bea
 &&\hat{p}_1-\hat{p}_2 = \hat{p}_3-\hat{p}_4 = 0, \;\;\;\;\; 
 (\mbox{mod $2\pi$}) \nonumber\\
 &&\hat{p}_1-\hat{p}_3 = \frac{2n\pi}{N}. \;\;\;\;\; 
 (n\in\mathbb{Z}, \;\;\; n\not\in N\mathbb{Z})
\eea
Solutions are  
\be
 (\hat{p}_1,\hat{p}_2,\hat{p}_3,\hat{p}_4) 
 = \brkt{\frac{n\pi}{N},\frac{n\pi}{N},-\frac{n\pi}{N},-\frac{n\pi}{N}}, 
\ee
where $n=1,\cdots,N-1$. 
Therefore the zero-mode conditions for $\bdm{(2,2)_{\pm 2}}$ are expressed as
\be
 \pm\frac{2n\pi}{N} = \frac{2\pi}{N}. 
\ee
As a result, the scalar zero-modes we have are 
\bea
 \bdm{(2,2)_{+2}}, \;\; \bdm{(2,2)_{-2}} &:& (\mbox{when $N=2$}) \nonumber\\
 \bdm{(2,2)_{+2}} &:& (\mbox{when $N=3,4,6$ and $n=1$}) \nonumber\\
 \bdm{(2,2)_{-2}} &:& (\mbox{when $N=3,4,6$ and $n=N-1$}) \nonumber\\
 \mbox{Nothing} &:& (\mbox{in the other cases}) 
\eea
}

\subsubsection{SO(7)} \label{zm:SO7}
The roots are $\{\pm\bdm{e}^i\pm\bdm{e}^j,\pm\bdm{e}^i\}$ ($1\leq i\neq j\leq 3$). 
Essentially inequivalent choices of 
the ${\rm SU(2)}_{\rm L}\times {\rm SU(2)}_{\rm R}$ subgroup are 
\be
 (\alp_L,\alp_R) = (\bdm{e}^1+\bdm{e}^2,\bdm{e}^1-\bdm{e}^2), \;\;
 (\bdm{e}^1+\bdm{e}^2,\bdm{e}^3), \;\;
 (\bdm{e}^3,\bdm{e}^1+\bdm{e}^2). 
\ee

\begin{description}
\item[(I) $\bdm{(\alp_L,\alp_R)=(e^1+e^2,e^1-e^2)}$] \mbox{}\\
The $U(1)_X$ generator is 
\be
 Q_X = \bdm{e}^3\cdot H = H_3. 
\ee
The irreducible decomposition of the adjoint representation of $G$ is 
\bea
 \bdm{21} &\bdm{=} & \bdm{(3,1)_0+(1,3)_0+(2,2)_{+1}+(2,2)_{-1}+(2,2)_0} \nonumber\\
 &&\bdm{+(1,1)_{+1}+(1,1)_{-1}+(1,1)_0},  \label{adj_decomp:SO7}
\eea
where $\bdm{(3,1)_0}$, $\bdm{(1,3)_0}$ and $\bdm{(1,1)_0}$ correspond to 
${\rm SU(2)}_{\rm L}$, ${\rm SU(2)}_{\rm R}$ and ${\rm U(1)}_{\rm X}$ generators, 
respectively. 
Thus candidates for the scalar zero-modes are three bidoublets and two singlets. 
Independent conditions in (\ref{zm_cond:gauge}) are expressed as 
\bea
 &&p_1+p_2 = p_1-p_2 = 0, \;\;\;\;\; (\mbox{mod $2\pi$}) \nonumber\\
 &&p_1+p_3,p_1,p_3 = \frac{2n_P\pi}{N}. \;\;\;\;\; 
 (n_P\in\mathbb{Z}, \;\;\; n_P\not\in N\mathbb{Z})
\eea
Solutions exist only when $N=4,6$, and they are 
\be
 P = \exp\brc{i\pi\brkt{H_1+H_2+\frac{2n_P}{N}H_3}}, 
\ee
where $n_P\neq 0,N/2$. 
Therefore the zero-mode conditions~(\ref{zm_cond:scalar}) 
for $\bdm{(2,2)_{\pm 1}}$, $\bdm{(2,2)_0}$ and $\bdm{(1,1)_{\pm 1}}$ are 
\be
 \pi\pm\frac{2n_P\pi}{N} = \frac{2\pi}{N}, \;\;\;\;\;
 \pi = \frac{2\pi}{N}, \;\;\;\;\;
 \pm\frac{2n_P\pi}{N} = \frac{2\pi}{N},  
\ee
respectively. 
Here the double signs correspond. 

When $N=4$, the scalar zero-modes we have are 
\bea
 \bdm{(2,2)_{-1}}, \;\; \bdm{(1,1)_{+1}} &:& (\mbox{when $n_P=1$}) \nonumber\\
 \bdm{(2,2)_{+1}}, \;\; \bdm{(1,1)_{-1}} &:& (\mbox{when $n_P=3$}) 
\eea

When $N=6$, they are 
\bea
 \bdm{(1,1)_{+1}} &:& (\mbox{when $n_P=1$}) \nonumber\\
 \bdm{(2,2)_{-1}} &:& (\mbox{when $n_P=2$}) \nonumber\\
 \bdm{(2,2)_{+1}} &:& (\mbox{when $n_P=4$}) \nonumber\\
 \bdm{(1,1)_{-1}} &:& (\mbox{when $n_P=5$})
\eea

\item[(II) $\bdm{(\alp_L,\alp_R)=(e^1+e^2,e^3)}$] \mbox{}\\
The $U(1)_X$ generator is 
\be
 Q_X = (\bdm{e}^1-\bdm{e}^2)\cdot H = H_1-H_2. 
\ee
The irreducible decomposition of the adjoint representation of $G$ is 
\bea
 \bdm{21} &\bdm{=}& \bdm{(3,1)_0+(1,3)_0+(2,3)_{+1}+(2,3)_{-1}} \nonumber\\
 &&\bdm{+(1,1)_{+2}+(1,1)_{-2}+(1,1)_0}. 
\eea
Candidates for the scalar zero-modes are $\bdm{(2,3)_{\pm 1}}$ 
and $\bdm{(1,1)_{\pm 1}}$. 
Independent conditions in (\ref{zm_cond:gauge}) are expressed as
\bea
 &&p_1+p_2 = p_3 = 0, \;\;\;\;\; (\mbox{mod $2\pi$}) \nonumber\\
 &&p_1+p_3, p_2+p_3, p_1-p_2 = \frac{2n_P\pi}{N}. \;\;\;\;\;
 (n_P\in\mathbb{Z}, \;\; n_P\not\in N\mathbb{Z}) 
\eea
Solutions exist when $N=3,4,6$, and they are  
\be
 P = \exp\brc{\frac{2n_P\pi i}{N}\brkt{H_1-H_2}}, 
\ee
where $n_P\neq 0,N/2$. 
Therefore the zero-mode conditions~(\ref{zm_cond:scalar}) for $\bdm{(2,3)_{\pm 1}}$ 
and $\bdm{(1,1)_{\pm 1}}$ are 
\be
 \pm\frac{2n_P\pi}{N} = \frac{2\pi}{N}, \;\;\;\;\;
 \pm\frac{4n_P\pi}{N} = \frac{2\pi}{N}, 
\ee
respectively. 

When $N=3$, the scalar zero-modes we have are 
\bea
 \bdm{(2,3)_{+1}}, \;\; \bdm{(1,1)_{-2}} &:& (\mbox{when $n_P=1$}) \nonumber\\
 \bdm{(2,3)_{-1}}, \;\; \bdm{(1,1)_{+2}} &:& (\mbox{when $n_P=2$}) 
\eea

When $N=4$, they are 
\bea
 \bdm{(2,3)_{+1}} &:& (\mbox{when $n_P=1$}) \nonumber\\
 \bdm{(2,3)_{-1}} &:& (\mbox{when $n_P=3$})
\eea

When $N=6$, they are 
\bea
 \bdm{(2,3)_{+1}} &:& (\mbox{when $n_P=1$}) \nonumber\\
 \mbox{Nothing} &:& (\mbox{when $n_P=2,4$}) \nonumber\\
 \bdm{(2,3)_{-1}} &:& (\mbox{when $n_P=5$}) 
\eea

\item[(III) $\bdm{(\alp_L,\alp_R)=(e^3,e^1+e^2)}$] \mbox{}\\
The results are obtained by exchanging ${\rm SU(2)}_{\rm L}$ and ${\rm SU(2)}_{\rm R}$ 
in the case (II). 
Hence we do not have ${\rm SU(2)}_{\rm L}$-doublet Higgses. 
\end{description}

\subsubsection{Sp(6)}
The roots are $\{\pm\bdm{e}^i\pm\bdm{e}^j,\pm 2\bdm{e}^i\}$ ($1\leq i\neq j\leq 3$). 
Essentially inequivalent choices of 
the ${\rm SU(2)}_{\rm L}\times {\rm SU(2)}_{\rm R}$ are 
\be
 (\alp_L,\alp_R) = (2\bdm{e}^1,2\bdm{e}^2), \;\; (\bdm{e}^1+\bdm{e}^2,2\bdm{e}^3), 
 \;\; (2\bdm{e}^3,\bdm{e}^1+\bdm{e}^2). 
\ee

\begin{description}
\item[(I) $\bdm{(\alp_L,\alp_R)=(2e^1,2e^2)}$] \mbox{}\\
The $U(1)_X$ generator is 
\be
 Q_X = \bdm{e}^3\cdot H = H_3. 
\ee
The irreducible decomposition of the adjoint representation of $G$ is 
\bea
 \bdm{21} &\bdm{=}& \bdm{(3,1)_0+(1,3)_0+(2,2)_0+(2,1)_{+1}+(2,1)_{-1}} \nonumber\\
 &&\bdm{+(1,2)_{+1}+(1,2)_{-1}+(1,1)_{+2}+(1,1)_{-2}+(1,1)_0}. 
\eea
Independent conditions in (\ref{zm_cond:gauge}) are expressed as
\bea
 &&2p_1 = 2p_2 = 0, \;\;\;\;\; (\mbox{mod $2\pi$}) \nonumber\\
 &&p_1+p_2, p_1\pm p_3, p_2\pm p_3, 2p_3 = \frac{2n_P\pi}{N}, 
 \;\;\;\;\; (n_P\in\mathbb{Z}, \;\; n_P\not\in N\mathbb{Z})
\eea
Solutions exist only when $N=4,6$. 
They are 
\be
 P = \begin{cases} {\displaystyle 
 P^{(1)}_{n_P} \equiv \exp\brc{i\pi\brkt{H_2+\frac{2n_P\pi}{N}H_3}}}, 
 & \\ 
 {\displaystyle 
 P^{(2)}_{n_P} \equiv \exp\brc{i\pi\brkt{H_1+\frac{2n_P\pi}{N}H_3}}}, \end{cases}
\ee
where $n_P\neq 0,N/2$. 

When $N=4$, the scalar zero-modes we have are 
\bea
 \bdm{(2,1)_{+1}}, \;\; \bdm{(1,2)_{-1}} &:& (\mbox{for $P_1^{(1)}$ or $P_3^{(2)}$})
 \nonumber\\
 \bdm{(2,1)_{-1}}, \;\; \bdm{(1,2)_{+1}} &:& (\mbox{for $P_3^{(1)}$ or $P_1^{(2)}$}) 
\eea

When $N=6$, they are 
\bea
 \bdm{(2,1)_{+1}} &:& (\mbox{for $P_1^{(1)}$ or $P_4^{(2)}$}) \nonumber\\
 \bdm{(1,2)_{-1}} &:& (\mbox{for $P_2^{(1)}$ or $P_5^{(2)}$}) \nonumber\\
 \bdm{(1,2)_{+1}} &:& (\mbox{for $P_4^{(1)}$ or $P_1^{(2)}$}) \nonumber\\
 \bdm{(2,1)_{-1}} &:& (\mbox{for $P_5^{(1)}$ or $P_2^{(2)}$}) 
\eea

\item[(II) $\bdm{(\alp_L,\alp_R)=(e^1+e^2,2e^3)}$] \mbox{}\\
The $U(1)_X$ generator is 
\be
 Q_X = \brkt{\bdm{e}^1-\bdm{e}^2}\cdot H = H_1-H_2. 
\ee
The irreducible decomposition of the adjoint representation of $G$ is 
\be
 \bdm{21 = (3,1)_0+(1,3)_0+(3,1)_{+2}+(3,1)_{-2}+(2,2)_{+1}+(2,2)_{-1}+(1,1)_0}. 
\ee
Independent conditions in (\ref{zm_cond:gauge}) are expressed as 
\bea
 &&p_1+p_2 = 2p_3 = 0, \;\;\;\;\; (\mbox{mod $2\pi$}) \nonumber\\
 &&p_1+p_3, p_2+p_3, 2p_1, 2p_2 = \frac{2n_P\pi}{N}, 
 \;\;\;\;\; (n_P\in\mathbb{Z}, \;\; n_P\not\in N\mathbb{Z})
\eea
where $n_P\neq 0,N/2$. 
Solutions exist only when $N=3,4,6$. 
They are 
\be
 P = \begin{cases} {\displaystyle 
 P_{n_P}^{(1)} \equiv \exp\brc{\frac{2n_P\pi i}{N}(H_1-H_2)}}, 
 & \\
 {\displaystyle 
 P_{n_P}^{(2)} \equiv \exp\brc{i\pi\brkt{\frac{2n_P-N}{N}\brkt{H_1-H_2}+H_3}}}. 
 \end{cases} 
\ee 

When $N=3$, the scalar zero-modes we have are 
\bea
 \bdm{(3,1)_{-2}}, \;\; \bdm{(2,2)_{+1}} &:& (\mbox{for $P_1^{(1)}$ or $P_1^{(2)}$}) 
 \nonumber\\
 \bdm{(3,1)_{+2}}, \;\; \bdm{(2,2)_{-1}} &:& (\mbox{for $P_2^{(1)}$ or $P_2^{(2)}$}) 
\eea

When $N=4$, they are
\bea
 \bdm{(2,2)_{+1}} &:& (\mbox{for $P_1^{(1)}$ or $P_1^{(2)}$}) \nonumber\\
 \bdm{(2,2)_{-1}} &:& (\mbox{for $P_3^{(1)}$ or $P_3^{(2)}$}) 
\eea

When $N=6$, they are 
\bea
 \bdm{(2,2)_{+1}} &:& (\mbox{for $P_1^{(1)}$ or $P_1^{(2)}$}) \nonumber\\
 \bdm{(2,2)_{-1}} &:& (\mbox{for $P_5^{(1)}$ or $P_5^{(2)}$}) \nonumber\\
 \mbox{Nothing} &:& (\mbox{in the other cases}) 
\eea

\item[(III) $\bdm{(\alp_L,\alp_R)=(2e^3,e^1+e^2)}$] \mbox{}\\
The results are obtained by exchanging ${\rm SU(2)}_{\rm L}$ and ${\rm SU(2)}_{\rm R}$ 
in the case (II). 
\end{description}

\section{Custodial symmetry and Weinberg angle} \label{CW}
\subsection{Custodial symmetry} \label{Custodial}
Here we consider a condition that the custodial symmetry 
is preserved after the electroweak symmetry is broken. 
The SU(2)${}_{\rm L}$ and SU(2)${}_{\rm R}$ generators are 
\be
 (T_L^\pm,T_L^3) = \brkt{\frac{E_{\pm\alp_L}}{\abs{\alp_L}},
 \frac{\alp_L\cdot H}{\abs{\alp_L}^2}}, \;\;\;\;\;
 (T_R^\pm,T_R^3) = \brkt{\frac{E_{\pm\alp_R}}{\abs{\alp_R}},
 \frac{\alp_R\cdot H}{\abs{\alp_R}^2}}, \label{def:T_LR}
\ee
respectively. 
Thus (\ref{decomp:A_M}) is rewritten as
\bea
 A_\mu \eql W_{L\mu}^+T_L^++W_{L\mu}^-T_L^-+W_{L\mu}^3T_L^3
 +W_{R\mu}^+T_R^++W_{R\mu}^-T_R^-+W_{R\mu}^3T_R^3 \nonumber\\
 &&+B_\mu^X {\rm x}\cdot H+\cdots,  \label{decomp:A_M2}
\eea
where 
\bea
 W_{L\mu}^\pm \defa \abs{\alp_L}W^{\pm\alp_L}_\mu, \;\;\;\;\;
 W_{L\mu}^3 \equiv \alp_L\cdot C_\mu, \nonumber\\
 W_{R\mu}^\pm \defa \abs{\alp_R}W^{\pm\alp_R}, \;\;\;\;\;
 W_{R\mu}^3 \equiv \alp_R\cdot C_\mu, 
\eea
and $B_\mu^X \equiv \frac{{\rm x}\cdot C_\mu}{\abs{\rm x}^2}$ is 
the ${\rm U(1)}_X$ gauge field that does not exist when $r=2$. 
The ellipsis denotes components that do not have zero-modes. 
Since the generators in (\ref{def:T_LR}) are normalized as
\be
 \tr\brkt{T_L^+T_L^-} = \tr\brkt{(T_L^3)^2} = \frac{1}{\abs{\alp_L}^2}, \;\;\;\;\;
 \tr\brkt{T_R^+T_R^-} = \tr\brkt{(T_R^3)^2} = \frac{1}{\abs{\alp_R}^2}, 
 \label{T_LR:normalize}
\ee
the canonically normalized zero-mode gauge fields are 
\be
 \hat{W}_{L\mu}^{\pm,3} \equiv \frac{\sqrt{\cA}}{g_A\abs{\alp_L}}W_{L\mu}^{\pm,3}, 
 \;\;\;\;\;
 \hat{W}_{R\mu}^{\pm,3} \equiv \frac{\sqrt{\cA}}{g_A\abs{\alp_R}}W_{R\mu}^{\pm,3}, 
 \;\;\;\;\;
 \hat{B}_\mu^Z \equiv \frac{\sqrt{\cA}}{g_Z}B_\mu^Z, \label{W_LR:normalize}
\ee
where $\cA$ is the area of the fundamental domain of $T^2/Z_N$. 

Since we have assumed that ${\rm SU(2)}_{\rm R}\times {\rm U(1)}_Z$ is unbroken by 
the orbifold boundary conditions, we introduce some 4D scalar fields 
at one of the fixed points of $T^2/Z_N$ 
in order to break it to ${\rm U(1)}_Y$. 
We demand that the custodial symmetry~${\rm SU(2)}_V\subset 
{\rm SU(2)}_{\rm L}\times {\rm SU(2)}_{\rm R}$ 
remains unbroken after the Higgs fields have VEVs. 
The generators of ${\rm SU(2)}_V$ are 
\bea
 T_V^\pm \defa T_L^\pm+T_R^\pm = \frac{E_{\pm\alp_L}}{\abs{\alp_L}}
 +\frac{E_{\pm\alp_R}}{\abs{\alp_R}}, \nonumber\\
 T_V^3 \defa T_L^3+T_R^3 = \frac{\alp_L\cdot H}{\abs{\alp_L}^2}
 +\frac{\alp_R\cdot H}{\abs{\alp_R}^2}. 
\eea
Thus the conditions for ${\rm SU(2)}_V$ to be unbroken are 
\bea
 \sbk{T_V^\pm,\vev{A_z}} \eql \sum_\bt\vev{W_z^\bt}
 \brkt{\frac{N_{\pm\alp_L,\bt}E_{\bt\pm\alp_L}}{\abs{\alp_L}}
 +\frac{N_{\pm\alp_R,\bt}E_{\bt\pm\alp_R}}{\abs{\alp_R}}} = 0, \nonumber\\
 \sbk{T_V^3,\vev{A_z}} \eql \sum_\bt\vev{W_z^\bt}
 \brkt{\frac{\alp_L\cdot\bt}{\abs{\alp_L}^2}
 +\frac{\alp_R\cdot\bt}{\abs{\alp_R}^2}}E_\bt = 0, \label{cond_custodial}
\eea
since $C_z^i$ do not have zero-modes and thus $\vev{C_z^i}=0$.

\subsubsection{Rank-two groups}
Let us first consider the rank-two groups. 
We introduce the following Lagrangian at $z=0$.\footnote{
Of course, $\cL_{\rm loc}$ can be localized at other fixed point. } 
\be
 \cL_{\rm loc} = \brc{-\cD_\mu\phi^\dagger\cD^\mu\phi-V(\phi)}\dlt(z), 
 \label{L_bd}
\ee
where $\phi$ is a complex scalar field belonging to $\bdm{(1,2)_{+1/2}}$ 
under ${\rm SU(2)}_{\rm L}\times {\rm SU(2)}_{\rm R}\times {\rm U(1)}_Z$, 
and $V(\phi)$ is a potential that force $\phi$ to have a nonvanishing VEV. 
After $\phi$ gets a VEV, ${\rm SU(2)}_{\rm R}\times {\rm U(1)}_Z$ 
is broken to ${\rm U(1)}_Y$, 
and the corresponding massless gauge field is expressed as 
\be
 \hat{B}_\mu^Y \equiv \sin\tht_Z \hat{W}_{R\mu}^3+\cos\tht_Z\hat{B}_\mu^Z, 
 \label{def:B^Y}
\ee 
where a mixing angle~$\tht_Z$ is determined by $\tan\tht_Z=g_Z/(g_A\abs{\alp_R})$. 
The hypercharge operator~$Y$ is identified as 
\be
 Y = T_R^3+Q_Z = \frac{\alp_R\cdot H}{\abs{\alp_R}^2}+Q_Z. 
 \label{def:Y}
\ee

After $W_z^\bt$ have nonvanishing VEVs, 
${\rm SU(2)}_{\rm L}\times {\rm U(1)}_Y$ is broken 
to the electromagnetic symmetry~${\rm U(1)}_{\rm em}$. 
Since $W_z^\bt$ is ${\rm U(1)}_Z$ neutral and 
only $U(1)_{\rm em}$ neutral $W_z^\bt$ can have nonvanishing VEVs, 
the root~$\bt$ must satisfy
\be
 \frac{\alp_L\cdot\bt}{\abs{\alp_L}^2}+\frac{\alp_R\cdot\bt}{\abs{\alp_R}^2} = 0,  
 \label{neutral_cond}
\ee
if $\vev{W_z^\bt}\neq 0$. 
Thus the second condition in (\ref{cond_custodial}) 
is automatically satisfied. 
The roots that satisfy (\ref{neutral_cond}) are 
$\pm\bdm{e}^2\in\bdm{(2,2)}$ in SO(5), 
and $\pm\brkt{\frac{\bdm{e}^1}{2}-\frac{\bdm{e}^2}{2\sqrt{3}}}
\in\bdm{(2,4)}$ in G${}_2$. 
Then, from the first condition in (\ref{cond_custodial}), we obtain a condition, 
\be
 \abs{\vev{W_z^{e^2}}} = \abs{\vev{W_z^{-e^2}}}, \;\;\;\;\;
 \vev{W_z^\bt} = 0, \;\; (\bt\neq\pm\bdm{e}^2) \label{rel:VEV}
\ee
for SO(5), while no nonvanishing VEV is allowed for G${}_2$.

\subsubsection{Rank-three groups}
Next consider the rank-three groups. 
Since the unbroken gauge symmetry by the orbifold conditions is 
${\rm SU(2)}_{\rm L}\times {\rm SU(2)}_{\rm R}\times {\rm U(1)}_X\times {\rm U(1)}_Z$, 
let us first assume that $\phi$ in (\ref{L_bd}) 
also has a nonzero ${\rm U(1)}_X$ charge 
in order to obtain ${\rm SU(2)}_{\rm L}\times {\rm U(1)}_Y$ at low energies. 
Then the ${\rm U(1)}_Y$ gauge field~$B_\mu^Y$ becomes a linear combination 
of $W_{R\mu}^3$, $B_\mu^X$ and $B_\mu^Z$, and the hypercharge is 
identified as
\be
 Y = T_R^3+Q_X+Q_Z = \frac{\alp_R\cdot H}{\abs{\alp_R}^2}+{\rm x}\cdot H+Q_Z. 
\ee
Thus the condition~(\ref{neutral_cond}) now becomes 
\be
 \frac{\alp_L\cdot\bt}{\abs{\alp_L}^2}+\frac{\alp_R\cdot\bt}{\abs{\alp_R}^2}
 +{\rm x}\cdot\bt = 0.   
\ee
From this and the second condition in (\ref{cond_custodial}), 
both (\ref{neutral_cond}) and ${\rm x}\cdot\bt=0$ must be satisfied 
if $\vev{W_z^\bt}\neq 0$. 
Such roots do not exist among the zero-modes 
listed in Sec.~\ref{ZM:rank3}. 
Therefore we introduce two complex scalar fields~$\phi_1$ and $\phi_2$ 
instead of $\phi$ on the fixed point, 
\be
 \cL_{\rm loc} = \brc{-\cD_\mu\phi_1^\dagger\cD^\mu\phi_1
 -\cD_\mu\phi_2^\dagger\cD^\mu\phi_2-V(\phi_1,\phi_2)}\dlt(z), 
\ee
where $\phi_1$ and $\phi_2$ are complex scalars belonging to 
$\bdm{(1,2)_{0,+1/2}}$ and $\bdm{(1,1)_{+1,0}}$ respectively 
under ${\rm SU(2)}_{\rm L}\times {\rm SU(2)}_{\rm R}
\times {\rm U(1)}_X\times {\rm U(1)}_Z$, 
and $V(\phi_1,\phi_2)$ is a potential for them. 
Since $\phi_1$ is neutral for ${\rm U(1)}_X$, 
the ${\rm U(1)}_Y$ gauge field~$B_\mu^Y$ is now independent of $B_\mu^X$. 
Hence the hypercharge is identified as (\ref{def:Y}). 
The ${\rm U(1)}_X$ charges are no longer relevant to the ${\rm U(1)}_Y$ and 
${\rm U(1)}_{\rm em}$ charges because ${\rm U(1)}_X$ is completely broken by 
a VEV of another scalar~$\phi_2$. 
Thus the ${\rm U(1)}_Y$ gauge field is given by (\ref{def:B^Y}). 
In this case, the ${\rm U(1)}_{\rm em}$ neutral condition becomes (\ref{neutral_cond}), 
which is consistent with the second condition in (\ref{cond_custodial}).  
As a result, possible nonvanishing VEVs are 
as follows. 
\bea
 &&\abs{\vev{W_z^{\pm(e^1-e^3)}}} = \abs{\vev{W_z^{\pm(e^2-e^4)}}} 
 \in \bdm{(2,2)_{\pm 2}} \;\;\;\;\; \mbox{in SU(4)}, \nonumber\\
 &&\abs{\vev{W_z^{\pm(e^2+e^3)}}} = \abs{\vev{W_z^{\pm(-e^2+e^3)}}} 
 \in \bdm{(2,2)_{\pm 1}}, \;\;\;\;\;
 \abs{\vev{W_z^{\pm e^3}}} \in \bdm{(1,1)_{\pm 1}} \;\;\;\;\; 
 \mbox{in SO(7) (I)},  \nonumber\\
 &&\abs{\vev{W_z^{\pm(e^1-e^3)}}} = \abs{\vev{W_z^{\pm(-e^2+e^3)}}}
 \in \bdm{(2,2)_{\pm 1}} \;\;\;\;\; 
 \mbox{in Sp(6) (II), Sp(6) (III)},  \label{rel:VEV2}
\eea
where the double signs correspond. 

In summary, fields that can have nonzero VEVs are 
the neutral components of a bidoublet~$\bdm{(2,2)}$ or a singlet~$\bdm{(1,1)}$. 
The above conditions indicate that a bidoublet~$\cH_a$ must have a VEV:
\be
 \vev{\cH_a} = \frac{1}{2}\begin{pmatrix} v_a & \\ & v_a \end{pmatrix}, \label{VEV:cH}
\ee
where $v_a>0$, if we redefine a phase of each field component appropriately.

\subsection{Weinberg angle and weak gauge boson masses}
In the approximation that the W and Z bosons have constant profiles 
over the extra dimensions, 
the 4D ${\rm SU(2)}_{\rm L}$ and ${\rm U(1)}_Y$ gauge coupling constants are 
read off from couplings to the matter zero-modes, and are identified as
\be
 g = \frac{g_A\abs{\alp_L}}{\sqrt{\cA}}, \;\;\;\;\;
 g' = \frac{g_Ag_Z\abs{\alp_R}}{\sqrt{\cA(g_A^2\abs{\alp_R}^2+g_Z^2)}}. 
 \label{def:4Dg}
\ee
Thus the Weinberg angle is calculated as
\be
 \tan^2\tht_W \equiv \frac{g'}{g} = \frac{g_Z^2\abs{\alp_R}^2}
 {\abs{\alp_L}^2(g_A^2\abs{\alp_R}^2+g_Z^2)}.  \label{expr:tanthtW}
\ee
We can obtain the experimental value~$\tan^2\tht_W\simeq 0.30$ 
by tuning the ratio~$g_Z/g_A$. 

Next we derive the expressions of the W and Z boson masses. 
From (\ref{W_LR:normalize}) and (\ref{def:B^Y}), 
the expression~(\ref{decomp:A_M2}) becomes 
\be
 A_\mu = W_{L\mu}^+T_L^++W_{L\mu}^-T_L^-+W_{L\mu}^3T_L^3
 +\sin\tht_ZB^Y_\mu T_R^3+\cdots, 
\ee
where $B_\mu^Y\equiv\frac{g_A\abs{\alp_R}}{\sqrt{\cA}}\hat{B}^Y_\mu$, 
after the breaking~${\rm SU(2)}_{\rm R}\times {\rm U(1)}_Z\to {\rm U(1)}_Y$. 
Then it follows that
\bea
 \sbk{A_\mu,\vev{A_z}} \eql \sum_\bt W_z^\bt\left\{
 W_{L,\mu}^+\frac{N_{\alp_L,\bt}}{\abs{\alp_L}}E_{\bt+\alp_L}
 +W_{L\mu}^-\frac{N_{-\alp_L,\bt}}{\abs{\alp_L}}E_{\bt-\alp_L} \right.\nonumber\\
 &&\left.\hspace{20mm}
 +\brkt{W_{L\mu}^3\frac{\alp_L\cdot\bt}{\abs{\alp_L}^2}
 +B_\mu^Y\sin\tht_Z\frac{\alp_R\cdot\bt}{\abs{\alp_R}^2}}E_\bt\right\}. 
 \label{A_mu:A_z}
\eea
From the results in the previous subsections, 
the only components that contribute to the W and Z boson masses 
are the neutral components of the bidoublets.  
Since the roots that form a bidoublet are expressed as
\be
 \brkt{\begin{array}{ccc} \gm_a+\alp_L & \stackrel{\alp_R}{\longrightarrow} 
 & \gm_a+\alp_L+\alp_R \\
 \uparrow_{\alp_L} && \uparrow_{\alp_L} \\
 \gm_a & \stackrel{\alp_R}{\longrightarrow} & \gm_a+\alp_R \end{array}}, 
\ee
where $a$ labels the bidoublets, 
(\ref{A_mu:A_z}) are rewritten as
\bea
 \sbk{A_\mu,\vev{A_z}} \eql \sum_a\left[
 \vev{W_z^{\gm_a+\alp_L}}\brc{\frac{e^{i\zeta}}{\sqrt{2}}W_{L\mu}^- E_{\gm_a}
 +\brkt{\frac{1}{2}W_{L\mu}^3-\frac{\sin\tht_Z}{2}B_\mu^Y}E_{\gm_a+\alp_L}} 
 \right.\nonumber\\
 &&\hspace{7mm}\left.
 +\vev{W_z^{\gm_a+\alp_R}}\brc{\frac{e^{i\eta}}{\sqrt{2}}W_{L\mu}^+ E_{\gm_a+\alp_L+\alp_R}
 -\brkt{\frac{1}{2}W_{L\mu}^3-\frac{\sin\tht_Z}{2}B_\mu^Y}E_{\gm_a+\alp_R}}
 \right],  \nonumber\\
\eea
where $\gm_a$ is the $T_L^3=T_R^3=-1/2$ component of 
the zero-mode bidoublets~$\cH_a$. 
We have used that $\abs{N_{-\alp_L,\gm_a+\alp_L}}^2=\abs{N_{\alp_L,\gm_a+\alp_R}}^2
=\abs{\alp_L}^2/2$, and $\zeta\equiv\arg(N_{-\alp_L,\gm_a+\alp_L})$ 
and $\eta\equiv\arg(N_{\alp_L,\gm_a+\alp_R})$. 
Thus the relevant terms in 6D Lagrangian are calculated as 
\bea
 \cL \eql -\frac{1}{4g_A^2}\tr(F^{(A)MN}F^{(A)}_{MN})+\cdots
 = -\frac{1}{2g_A^2\pi^2 R_1^2}\tr\brkt{\sbk{A^\mu,\vev{A_z}}
 \sbk{A_\mu,\vev{A_z}}^\dagger}+\cdots \nonumber\\
 \eql -\sum_a\frac{\abs{\vev{W_z^{\gm_a+\alp_L}}}^2
 +\abs{\vev{W_z^{\gm_a+\alp_R}}}^2}
 {2g_A^2\pi^2R_1^2}\brc{\frac{1}{2}W_L^{+\mu}W_{L\mu}^-
 +\brkt{\frac{1}{2}W_{L\mu}^3-\frac{\sin\tht_Z}{2}B_\mu^Y}^2}+\cdots \nonumber\\
 \eql  -\frac{g^2\sum_a v_a^2}{4\cA}
 \brc{\hat{W}_L^{+\mu}\hat{W}_{L\mu}^-
 +\frac{1}{2}\brkt{\hat{W}_{L\mu}^3-\frac{\abs{\alp_R}\sin\tht_Z}{\abs{\alp_L}}
 \hat{B}_\mu^Y}^2}+\cdots.  \label{rel:WZmass}
\eea
At the last step, we have used that (\ref{W_LR:normalize}), and 
$\abs{\vev{W_z^{\gm_a+\alp_L}}}=\abs{\vev{W_z^{\gm_a+\alp_R}}}
\equiv g\pi R_1 v_a/\sqrt{2}\abs{\alp_L}$ 
($g$: 4D ${\rm SU(2)}_{\rm L}$ gauge coupling),  
which follows from (\ref{rel:VEV}) or (\ref{rel:VEV2}). 
We obtain the W and Z boson mass terms 
by integrating (\ref{rel:WZmass}) over the extra dimensions, 
and their masses are read off as 
\bea
 m_W \eql \frac{g}{2}\sqrt{\sum_a v_a^2}, \nonumber\\
 m_Z \eql \brkt{1+\frac{\abs{\alp_R}^2\sin^2\tht_Z}{\abs{\alp_L}^2}}^{1/2}m_W 
 = \brkt{1+\frac{g_Z^2\abs{\alp_R}^2}{\abs{\alp_L}^2(g_A^2\abs{\alp_R}^2+g_Z^2)}}^{1/2}
 m_W.  \label{expr:m_W}
\eea
From these and (\ref{expr:tanthtW}), we find that 
$\rho\equiv m_W^2/(m_Z^2\cos^2\tht_W)=1$. 
This is expected because we have assumed that only ${\rm SU(2)}_{\rm L}$ doublets 
and singlets have nonzero VEVs and neglected the $z$-dependence of the mode functions 
for the W and Z bosons. 
The custodial symmetry will play a crucial role 
when such $z$-dependence is taken into account.

\ignore{
\section{Matter fields}
In this section, we consider the matter fermionic fields. 
There are two possibilities according to where they live. 
In the following, we focus on the quark sector, 
but a similar argument is also applicable to the lepton sector. 
\subsection{Localized fermions}
The quarks~($q_{Li},u_{Ri},d_{Ri}$) 
and leptons~($l_{Li},e_{Ri},\nu_{Ri}$), where $i=1,2,3$ denotes the generation, 
can be localized at a fixed point of $T^2/Z_N$. 
In this case, they are embedded into multiplets of the residual gauge symmetry 
under the orbofold conditions. 
In the previous section, 
we have seen that the Higgs fields that break the electroweak symmetry 
belong to $\bdm{(2,2)_0}$ for 
${\rm SU(2)}_{\rm L}\times {\rm SU(2)}_{\rm R}\times {\rm U(1)}_Z$ in all possible cases. 
Thus the embedding of the quark and lepton fields is restricted 
from a condition that the Yukawa couplings 
relevant to the quark and lepton masses are allowed 
by the residual symmetry.  
In the following, we focus on the quark sector. 
Similar embeddings are also possible for the lepton sector. 
A possible embedding into 
${\rm SU(2)}_{\rm L}\times {\rm SU(2)}_{\rm R}\times {\rm U(1)}_Z$ 
representations is as follows. 
The right-handed quarks are embedded as
\be
 u_{Ri} \in \bdm{(1,1)_{2/3}}, \;\;\;\;\;
 d_{Ri} \in \bdm{(1,1)_{-1/3}},  \label{embed:qR}
%
%
\ee
and the left-handed fermions are bidoublets. 
We omit the $U(1)_X$ charges because they are irrelevant to the hypercharge. 
Since the Higgs field~$\cH$ belongs to $\bdm{(2,2)_0}$, 
the following two 
left-handed bidoublet fermions are necessary to form the Yukawa couplings. 
\be
 \cQ_{Li}^{(2/3)} \equiv (Q_{Li}^{(1)},Q_{Li}^{(2)}) \in \bdm{(2,2)_{2/3}}, \;\;\;\;\;
 \cQ_{Li}^{(-1/3)} \equiv (Q_{Li}^{(3)},Q_{Li}^{(4)}) \in \bdm{(2,2)_{-1/3}},  
 \label{embed:qL}
%
%
\ee
where $\{Q_{Li}^{(1)},Q_{Li}^{(3)}\}$ 
and $\{Q_{Li}^{(2)},Q_{Li}^{(4)}\}$ 
are ${\rm SU(2)}_{\rm L}$ doublets whose $T_R^3$ eigenvalues 
are $-1/2$ and $1/2$, respectively. 
The Higgs bidoublet is decomposed as $\cH=(\tl{H}_2,H_1)$, 
where $\tl{H}_2^i\equiv\ep_{ij}H_2^{j*}$ and $H_1^i$ are ${\rm SU(2)}_{\rm L}$ doublets 
whose $T_R^3$ eigenvalues are $-1/2$ and $1/2$, respectively. 
Another possible embedding is 
\bea
 u_{Ri} \in \bdm{(1,1)_{2/3}}, \;\;\;\;\;
 d_{Ri} \in \bdm{(1,3)_{2/3}}, \;\;\;\;\;
 q_{Li} \in \bdm{(2,2)_{2/3}}. 
\eea
In this case, only one bidoublet fermion is necessary for each left-handed quark~$q_{Li}$. 
The right-handed quark~$d_{Ri}$ is the $T_R^3=-1$ component of $\bdm{(1,3)_{2/3}}$. 
Note that we cannot identify the $T_R^3=0$ component of 
the same multiplet as $u_{Ri}$ because the up-type and down-type quarks would have 
a degenerate mass in that case. 
Embeddings into higher dimensional representations of ${\rm SU(2)}_{\rm R}$ are also 
possible,\footnote{
An embedding~$(u_{Ri},d_{Ri})\in\bdm{(1,2)_{1/6}}$ and $q_{Li}\in\bdm{(2,1)_{1/6}}$  
is not allowed because the masses of the up-type and down type quarks 
would be degenerate. 
}  
but more exotic particles appear in such cases and we need more 4D localized 
fermions in order to decouple them. 
}

\section{Matter field} \label{bulk_fermion}
We consider a case that quarks and leptons live in the bulk.  
This case is interesting 
because the hierarchical structure of the Yukawa coupling constants 
can be realized by the wave function localization~\cite{ArkaniHamed:1999dc,Kaplan:2000av}, 
and the generation structure can also be obtained 
by a background magnetic flux~\cite{Abe:2013bca}. 
In the following, we focus on the quark sector, 
but a similar argument is also applicable to the lepton sector. 

\subsection{Zero-mode condition}
A 6D Weyl fermion~$\Psi_{\chi_6}$ with the 6D chirality~$\chi_6=\pm$ is decomposed as
\be
 \Psi_{\chi_6} = \sum_{\chi_4=\pm}\Psi_{\chi_6,\chi_4}, 
\ee
where $\chi_4=+(R),-(L)$ is the 4D chirality. 
The orbifold boundary conditions for $\Psi_{\chi_6,\chi_4}$ 
are given by~\cite{Scrucca:2003ut} 
\bea
 \Psi_{\chi_6,\chi_4}(x,z+1) \eql \Psi_{\chi_6,\chi_4}(x,z), 
 \nonumber\\
 \Psi_{\chi_6,\chi_4}(x,z+\tau) \eql \Psi_{\chi_6,\chi_4}(x,z), 
 \nonumber\\
 \Psi_{\chi_6,\chi_4}(x,\omg z) \eql \omg^{-\frac{\chi_4\chi_6}{2}}
 e^{i\vph_\omg}P\Psi_{\chi_6,\chi_4}(x,z).  \label{ob_cond:fermion}
\eea
A factor~$\omg^{-\frac{\chi_4\chi_6}{2}}$ appears 
because a 6D spinor is charged under 
a rotation in the extra-dimensional space. 
The phase~$\vph_\omg$ satisfies (\ref{cond_for_vphomg}). 

As pointed out in Ref.~\cite{Abe:2009uz}, 
the generations and the hierarchy among the Yukawa couplings 
can be obtained by introducing an extra gauge symmetry~$G_F$ and 
assuming a magnetic flux on $T^2/Z_N$ and the Wilson line phases for $G_F$.  
The zero-modes are contained in $\Psi_{\chi_6,\chi_4}$ as 
\be
 \Psi_{\chi_6,\chi_4}(x,z) = \sum_{j=1}^{j_{\rm max}}
 \sum_{\mu} f_{\chi_6}^{(j)\mu}(z)|\mu\rangle\psi_{\chi_4}^{(j)\mu}(x)+\cdots, 
 \label{KKdecomp:fermion2}
\ee
where $\mu$ runs over the weights of the zero-mode states,\footnote{
Do not confuse it with the 4D Lorentz index. }  
and the ellipsis denotes the nonzero KK modes. 
The number of the zero-modes~$j_{\rm max}$ is determined by 
the magnetic flux~\cite{Abe:2013bca}. 
The zero-mode functions~$f_{\chi_6}^{(j)\mu}(z)$ are determined so that 
(\ref{KKdecomp:fermion2}) satisfies the first two conditions in (\ref{ob_cond:fermion}). 
From the last condition in (\ref{ob_cond:fermion}), we obtain
\be
 \psi_{\chi_4}^{(j)\mu}(x) = \omg^{-\frac{\chi_4\chi_6}{2}}e^{i\vph_\omg}P
 \psi_{\chi_4}^{(j)\mu}(x). 
\ee
Namely, the zero-mode is an eigenvector of 
$\omg^{-\frac{\chi_4\chi_6}{2}}e^{i\vph_\omg}P$ with an eigenvalue 1. 
Denote the highest weight of a representation~$\cR$ 
that $\Psi_{\chi_4,\chi_6}$ belongs to as $\mu_{\rm max}$. 
Then $\mu$ is expressed as
\be
 \mu = \mu_{\rm max}-\sum_i k_i\alp_i, 
\ee
where $k_i$ are non-negative integers, and $\alp_i$ are the simple roots. 
Since $P^N|\mu\rangle = e^{iNp\cdot\mu}|\mu\rangle 
= e^{iNp\cdot\mu_{\rm max}}|\mu\rangle$,\footnote{
We have used (\ref{value:palp}) at the second equality. 
} 
the phase~$\vph_\omg$ is determined by (\ref{cond_for_vphomg}) as 
$\vph_\omg=\frac{\pi}{N}(2m_\omg+1)-p\cdot\mu_{\rm max}$, 
where $m_\omg=0,1,\cdots,N-1$. 
Thus we find that 
\bea
 \omg^{-\frac{\chi_4\chi_6}{2}}e^{i\vph_\omg}P|\mu\rangle 
 \eql e^{-\frac{2\pi i}{N}\cdot\frac{\chi_4\chi_6}{2}}
 \exp\brkt{\frac{(2m_\omg+1)\pi i}{N}-ip\cdot\mu_{\rm max}}e^{ip\cdot\mu}
 |\mu\rangle \nonumber\\
 \eql \exp\brkt{\frac{\pi i(2m_\omg+1-\chi_4\chi_6)}{N}
 -i\sum_i k_i(p\cdot\alp_i)}|\mu\rangle. 
\eea
Namely, the zero-mode condition for the state~$|\mu\rangle$ is 
\be
 \frac{\pi(2m_\omg+1-\chi_4\chi_6)}{N}-\sum_ik_i(p\cdot\alp_i) = 0. 
 \;\;\;\;\; (\mbox{mod $2\pi$})
 \label{zm_cond:fermion}
\ee

\subsection{$\bdm{Zb_L\bar{b}_L}$ coupling}
When the quarks live in the bulk, 
the $Zb_L\bar{b}_L$ coupling often receives a large correction 
induced by mixing with the KK modes. 
The authors of Ref.~\cite{Agashe:2006at} pointed out that 
the custodial symmetry plays an important role 
to suppress the deviation of this coupling from 
the standard model value. 
The $Zb_L\bar{b}_L$ coupling is protected if the theory has 
a parity symmetry~$\cP_{\rm LR}$ that exchanges 
${\rm SU(2)}_{\rm L}$ and ${\rm SU(2)}_{\rm R}$,  
and $b_L$ is the component of $T_L^3=T_R^3=-\frac{1}{2}$ 
in a bidoublet~$\bdm{(2,2)}$. 
Since the Higgs fields also belong to $\bdm{(2,2)}$, the right-handed quarks should 
belong to $\bdm{(1,1)}$ or $\bdm{(1,3)+(3,1)}$. 

Cases in which the bosonic sector has the parity symmetry~$\cP_{\rm LR}$ 
and a scalar bidoublet 
are SO(5), SU(4) and SO(7) (I) in Sec.~\ref{ZM:gauge}. 
In Appendix~\ref{decompList}, 
we list the irreducible representations of these groups 
whose dimensions are less than 30, and their decomposition 
into the ${\rm SU(2)}_{\rm L}\times {\rm SU(2)}_{\rm R}
(\times {\rm U(1)}_X)$ multiplets. 
There is no $\bdm{(1,3)+(3,1)}$ multiplets included in the list. 
Hence the left-handed and the right-handed quarks 
should be embedded into $\bdm{(2,2)}$ and $\bdm{(1,1)}$, respectively.

\subsection{Yukawa couplings}
\subsubsection{General expression}
The Yukawa couplings originate from the 6D minimal couplings 
in the kinetic term, $i\bar{\Psi}_{\chi_6}\Gm^M\cD_M\Psi_{\chi_6}
=-\frac{i\chi_6}{\pi R_1}\bar{\Psi}_{\chi_6,\chi_4=\chi_6}A_z\Psi_{\chi_6,\chi_4=-\chi_6}
+\hc+\cdots$. 
\ignore{
Let us factor out the group-theoretical factor 
from the zero-modes in (\ref{KKdecomp:fermion2}) as 
\bea
 \Psi_{\chi_6}(x,z) \eql \dlt_{\chi_6,\sgn(M)}
 \sum_{j=1}^{3}\left\{
 \sum_{\cR_R}f_{\chi_6,0}^{(j)\cR_R}(z)|\mu_R\rangle\psi_R^{(j)\cR_R}(x)
 +\sum_{\cR_L}f_{\chi_6,0}^{(j)\cR_L}(z)|\mu_L\rangle\psi_L^{(j)\cR_L}(x)\right\} 
 \nonumber\\
 &&+\cdots, 
\eea
where $\mu_{R,L}$ are the weights of the zero-mode states 
that belong to the irreducible representations~$\cR_{R,L}$ 
of ${\rm SU(2)}_{\rm L}\times{\rm SU(2)}_{\rm R}(\times{\rm U(1)}_X)\times{\rm U(1)}_Z$. }
The canonically normalized Higgs zero-mode~$H^\bt$ is contained in $A_z$ as
$A_z=\sum_\bt\frac{\sqrt{2}}{\abs{\alp_L}}g\pi R_1 H^\bt E_\bt+\cdots$, 
where $g$ is the ${\rm SU(2)}_{\rm L}$ 
gauge coupling constant. (See (\ref{com:AzAbz}).)
Then the Yukawa couplings in 4D effective Lagrangian 
are expressed as
\be
 \cL_{\rm yukawa} = \begin{cases} {\displaystyle \sum_{i,j}\left(
 \sum_{\bt,\mu_L}
 y_{(+)ij}^{\cR_H\cR_L\cR_R} \brkt{H^\bt}^*
 \bar{\psi}_L^{(i)\mu_L}\psi_R^{(j)\mu_L+\bt}+\hc \right)} 
 & (\chi_6=+) \\
 {\displaystyle \sum_{i,j}\left(\sum_{\bt,\mu_L}
 y_{(-)ij}^{\cR_H\cR_L\cR_R} H^\bt
 \bar{\psi}_L^{(i)\mu_L}\psi_R^{(j)\mu_L-\bt}+\hc \right)}
 & (\chi_6=-) \end{cases},  
 \label{L_yukawa:bulk}
\ee
where $\cR_H$, $\cR_L$ and $\cR_R$ are irreducible representations 
of ${\rm SU(2)}_{\rm L}\times{\rm SU(2)}_{\rm R}(\times{\rm U(1)}_X)\times{\rm U(1)}_Z$ 
that $|\bt\rangle$, $|\mu_L\rangle$ and 
$|\mu_R\rangle=|\mu_L+\chi_6\bt\rangle$ belong to, and 
\bea
 y_{(+)ij}^{\cR_H\cR_L\cR_R} \defa i\sqrt{2}g\langle\mu_L|E_{-\bt}|\mu_L+\bt\rangle
 \int\dr^2z\;f_{+,0}^{(i)\mu_L *}(z)f_{+,0}^{(j)\mu_L+\bt}(z), \nonumber\\
 y_{(-)ij}^{\cR_H\cR_L\cR_R} \defa i\sqrt{2}g\langle\mu_L|E_\bt|\mu_L-\bt\rangle
 \int\dr^2z\;f_{-,0}^{(i)\mu_L *}(z)f_{-,0}^{(j)\mu_L-\bt}(z). 
 \label{expr:yukawa_cp}
\eea
Note that these coupling constants only depend on 
the representations~$\{\cR_H,\cR_L,\cR_R\}$, 
and take common values for all $\bt\in\cR_H$ and $\mu_L\in\cR_L$. 

Exponentially small Yukawa couplings can be obtained 
by using the wave function localization 
in the extra dimensions~\cite{ArkaniHamed:1999dc,Kaplan:2000av}.
For the third generation, we assume that the overlap integrals 
in (\ref{expr:yukawa_cp}) do not provide 
any suppression factors, \ie, equal one. 
Then the Yukawa couplings are determined only by 
the group-theoretical factors. 
In the following, we focus on the third generation quarks. 

Consider a 6D Dirac fermion~$\Psi=\Psi_++\Psi_-$  
that belongs to the representation~$\cR$. 
The theory is assumed to be symmetric under an exchange:~$\Psi_+\leftrightarrow -\Psi_-$ 
so that a 6D mass term~$M_\Psi(\bar{\Psi}_+\Psi_-+\bar{\Psi}_-\Psi_+)$ 
is prohibited. 
Let us also assume that $\Psi_{\chi_6,-}$ and $\Psi_{\chi_6,+}$ 
have zero-modes $\cQ_L^{(\chi_6)}\in\bdm{(2,2)}$ 
and $\lmd_R^{(\chi_6)}\in\bdm{(1,1)}$. 
The Higgs fields~$H^\bt$ that couple to them form bidoublets~$\cH_a$. 
Then, from (\ref{L_yukawa:bulk}), the Yukawa couplings from 
$i\sum_{\chi_6=\pm}\bar{\Psi}_{\chi_6}\Gm^M\cD_M\Psi_{\chi_6}$ 
before the breaking of ${\rm SU(2)}_{\rm R}\times {\rm U(1)}_Z$ at the fixed point 
are expressed as 
\be
 \cL_{\rm yukawa} = \sum_a\brc{y^{(+)}_a\tr\brkt{\bar{\cQ}_L^{(+)}\tl{\cH}_a}\lmd_R^{(+)}
 +y^{(-)}_a\tr\brkt{\bar{\cQ}_L^{(-)}\cH_a}\lmd_R^{(-)}+\hc},  
 \label{expr:yukawa_cp3}
\ee
where $\tl{\cH}_a\equiv\sgm_2\cH_a^*\sgm_2$ and 
\bea
 y_a^{(+)} \eql i\sqrt{2}g
 \langle \mu_L|E_{-\bt}|\mu_L+\bt\rangle = i\sqrt{2}g N_{\bt,\mu_L}^*, \nonumber\\
 y_a^{(-)} \eql i\sqrt{2}g
 \langle \nu_L|E_\bt|\nu_L-\bt\rangle = i\sqrt{2}g N_{-\bt,\nu_L}^*. 
 \label{expr:yukawa_cp2}
\eea
Here $|\mu_L\rangle,|\nu_L\rangle\in\bdm{(2,2)}$, 
$|\mu_L+\bt\rangle,|\nu_L-\bt\rangle\in\bdm{(1,1)}$, and  
a complex constant~$N_{\bt,\mu}$ is defined below (\ref{master_formula}). 
Note that $\cQ_L^{(+)}$ and $\cQ_L^{(-)}$ ($\lmd_R^{(+)}$ and $\lmd_R^{(-)}$) 
belong to different $\bdm{(2,2)}$ ($\bdm{(1,1)}$) multiplets in $\cR$
because the same $\bdm{(2,2)}$ ($\bdm{(1,1)}$) cannot satisfy 
(\ref{zm_cond:fermion}) for $\chi_6=\pm$ simultaneously. 
We discriminate the two different $\bdm{(2,2)}$ and $\bdm{(1,1)}$ multiplets 
by denoting them as $\cQ_L^{(\chi_6)}\in\bdm{(2,2)_{\chi_6}}$ 
and $\lmd_R^{(\chi_6)}\in\bdm{(1,1)_{\chi_6}}$. 
The Yukawa couplings depend on how the quark fields are embedded 
into $\cQ_L^{(\pm)}$ and $\lmd_R^{(\pm)}$.

\subsubsection{Embedding of quarks} \label{embed_quark}
As we will see in Sec.~\ref{HiggsPotential}, 
the Higgs potential at tree level only contains quartic terms. 
The electroweak symmetry breaking occurs at one-loop level, 
and the top Yukawa coupling provides a dominant contribution to 
the one-loop Higgs potential. 
In general, such one-loop potential 
breaks ${\rm SU(2)}_{\rm L}\times {\rm SU(2)}_{\rm R}$, 
and thus the Higgs VEVs are not aligned as (\ref{VEV:cH}). 
Namely the custodial symmetry is broken. 
A simple way to avoid this difficulty is to assume that 
the quark fields couple to the Higgs fields 
only through a combination~$\cH_a+\tl{\cH}_a$. 
This is achieved when $y_a^{(+)}=y_a^{(-)*}$ 
and the quark fields are equally contained 
in both $\Psi_+$ and $\Psi_-$. 
Specifically, consider a case that $\nu_L=-\mu_L$ and 
4D fermions~$\zeta_R\in\bdm{(2,2)}$ and $\eta_L\in\bdm{(1,1)}$ 
are localized at a fixed point, which transform as $\zeta_R\to -\zeta_R$ 
and $\eta_L\to -\eta_L$ under $\Psi_\pm\to -\Psi_\mp$. 
Then combinations~$\cQ_L'\equiv(-\cQ_L^{(+)}+\cQ_L^{(-)})/\sqrt{2}$ 
and $\lmd_R'\equiv (-\lmd_R^{(+)}+\lmd_R^{(-)})/\sqrt{2}$ have masses with them 
at the fixed point and are decoupled at low energies. 
Since $y_a^{(+)}=y_a^{(-)*}$ due to the property~(\ref{rel:Ns}), 
we can redefine the overall phases of $\cH_a$ so that $y_a^{(+)}=y_a^{(-)}>0$. 
Then we obtain the desired form of the Yukawa coupling,\footnote{
Notice that $\cQ_R^{(\mp)}$ and $\lmd_L^{(\mp)}$ also satisfy  
the zero-mode condition~(\ref{zm_cond:fermion}) 
when $\cQ_L^{(\pm)}$ and $\lmd_R^{(\pm)}$ are zero-modes. 
So we also need additional 4D localized fermions to decouple them. 
} 
\be
 \cL_{\rm yukawa} = \frac{y_\lmd}{2}\sum_a\tr\brc{\bar{\cQ}_L
 \brkt{\cH_a+\tl{\cH}_a}}\lmd_R+\hc+\cdots, 
 \label{cL_yukawa}
\ee
where $y_\lmd\equiv y^{(+)}_a=y^{(-)}_a$, 
$\cQ_L\equiv(\cQ_L^{(+)}+\cQ_L^{(-)})/\sqrt{2}$ 
and $\lmd_R\equiv(\lmd_R^{(+)}+\lmd_R^{(-)})/\sqrt{2}$. 

Now we will see how the quark fields should be embedded into 6D fields. 
For simplicity, we consider a case that there is one Higgs bidoublet~$\cH$ 
as a zero-mode for a while. 
We introduce two 6D Dirac fermions~$\Psi^{(2/3)}=\Psi_+^{(2/3)}+\Psi_-^{(2/3)}$ 
and $\Psi^{(-1/3)}=\Psi_+^{(-1/3)}+\Psi_-^{(-1/3)}$, whose ${\rm U(1)}_Z$ charges 
are $2/3$ and $-1/3$, respectively. 
Let us assume that $\Psi^{(q_Z)}$ ($q_Z=2/3,-1/3$) 
contain $\cQ_L^{(q_Z)}\in\bdm{(2,2)}$ and $\lmd_R^{(q_Z)}\in\bdm{(1,1)}$ 
as zero-modes. 
The bidoublets are decomposed as 
\be
 \cQ_L^{(2/3)} = (Q_L^{(1)},Q_L^{(2)}), \;\;\;\;\;
 \cQ_L^{(-1/3)} = (Q_L^{(3)},Q_L^{(4)}), \;\;\;\;\;
 \cH = (\tl{H}_{2},H_{1}), 
\ee
where $\tl{H}_{2}^i\equiv\ep_{ij}H_2^{j*}$, and 
$\{Q_L^{(1)},Q_L^{(3)},\tl{H}_{2}\}$ and $\{Q_L^{(2)},Q_L^{(4)},H_{1}\}$ 
are ${\rm SU(2)}_{\rm L}$ doublets whose $T_R^3$ eigenvalues are $-1/2$ and $1/2$, 
respectively. 
Then the Yukawa couplings in the form of (\ref{cL_yukawa}) are expressed as 
\bea
 \cL_{\rm yukawa} \eql \frac{y_t}{2}\tr\brc{\bar{\cQ}_L^{(2/3)}\brkt{\cH+\tl{\cH}}}t_R
 +\frac{y_b}{2}\tr\brc{\bar{\cQ}_L^{(-1/3)}\brkt{\cH+\tl{\cH}}}b_R+\hc 
 \nonumber\\
 \eql \frac{y_t}{2}\brc{\bar{Q}_L^{(1)}\brkt{\tl{H}_2+\tl{H}_1}
 +\bar{Q}_L^{(2)}\brkt{H_1+H_2}}t_R \nonumber\\
 &&+\frac{y_b}{2}\brc{\bar{Q}_L^{(3)}\brkt{\tl{H}_2+\tl{H}_1}
 +\bar{Q}_L^{(4)}\brkt{H_1+H_2}}b_R+\hc, \label{L_yukawa:brane}
\eea
where $y_t$ and $y_b$ are calculated from (\ref{expr:yukawa_cp2}). 
Since only the combination~$H_1+H_2$ couples to the quarks,  
this combination obtains a tachyonic mass 
while the other combination~$H_1-H_2$ does not at one-loop level. 
Therefore the latter does not have a nonzero VEV, and $\vev{H_1}=\vev{H_2}$ is realized. 
Namely, the alignment~(\ref{VEV:cH}) is achieved. 
(See Sec.~\ref{Vtree:rank3}.)
Since $Q^{(1)}_{L}$ and $Q^{(4)}_{L}$ have the same quantum numbers 
for ${\rm SU(2)}_{\rm L}\times {\rm U(1)}_Y$, they are mixed with each other 
after the breaking~${\rm SU(2)}_{\rm R}\times {\rm U(1)}_Z\to {\rm U(1)}_Y$ occurs 
at the fixed point. 
The left-handed quark is identified as a linear combination, 
\be
 q_L = \cos\tht_{q}Q_{L}^{(1)}+\sin\tht_{q}Q_{L}^{(4)}, \label{ql:mixing}
\ee
where $\tht_{q}$ is a mixing angle.  
The orthogonal combination and $Q_{L}^{(2)}$ and $Q_{L}^{(3)}$ 
are exotic fields that must be decoupled at low energies. 
Hence we need to introduce 4D localized fermions 
that couple with those exotic components. 
As a result, the following Yukawa couplings are obtained at low energies.  
\be
 \cL_{\rm yukawa}^{SU(2)_L\times U(1)_Y} 
 = \frac{y_t}{2}\cos\tht_{q}q_{L}^\dagger\brkt{\tl{H}_2+\tl{H}_1}t_R
 +\frac{y_b}{2}\sin\tht_{q}q_{L}^\dagger\brkt{H_1+H_2}b_R+\hc. 
 \label{cL_yukawa2}
\ee
When $y_t=y_b$, the large ratio of the top quark mass~$m_t$ 
to the bottom quark mass~$m_b$ is obtained if $\tht_q=\cO(m_b/m_t)$.\footnote{
In contrast to the mixing between $\cQ_L^{(+)}$ and $\cQ_L^{(-)}$, 
the mixing angle~$\tht_q$ can take arbitrary values 
because there is no symmetry to fix it. 
} 
In such a case, $m_t$ is calculated as
\be
 m_t = \abs{\frac{y_t}{2}v\cos\tht_q} \simeq \abs{\frac{y_tv}{2}} 
 = \frac{g\abs{N_{\bt,\mu_L}}v}{\sqrt{2}} = \sqrt{2}\abs{N_{\bt,\mu_L}}m_W, 
\ee
where $v$ is defined as $\vev{H_1}=\vev{H_2}=(0,v/2)^t$ (see (\ref{VEV:cH})). 
We have used that $\cos\tht_q\simeq 1$, (\ref{expr:m_W}) and (\ref{expr:yukawa_cp2}).
Therefore, the observed top quark mass is obtained 
if $\abs{N_{\bt,\mu_L}}=\sqrt{2}$.\footnote{
A small deviation from the observed value of $m_t$ is expected to be 
explained by quantum correction. 
(See Ref.~\cite{Lim:2009mm}.)
}
We can extend this result to the two-Higgs-bidoublet case 
straightforwardly.

\subsubsection{Available representations for matter fermions} \label{avl_rep}
In summary, the quark multiplets should be embedded into 
two 6D Dirac fermions~$\Psi^{(2/3)}$ and $\Psi^{(-1/3)}$ 
whose ${\rm U(1)}_Z$ charges are $2/3$ and $-1/3$, respectively. 
Irreducible representations~$\cR$ which they belong to 
must satisfy the following conditions. 
\begin{enumerate}
\item $\cR$ includes two bidoublets and two singlets, 
which are denoted as $\bdm{(2,2)_\pm}$ and $\bdm{(1,1)_\pm}$, 
respectively. \label{condition1}

\item There are weights~$\mu_L$ and $\nu_L=-\mu_L$ 
that satisfy $|\mu_L\rangle\in\bdm{(2,2)_+}$, 
$|\mu_L+\bt\rangle\in\bdm{(1,1)_+}$, 
$|\nu_L\rangle\in\bdm{(2,2)_-}$, $|\nu_L-\bt\rangle\in\bdm{(1,1)_-}$, and 
$\abs{N_{\bt,\mu_L}} = \abs{N_{-\bt,\nu_L}} = \sqrt{2}$, 
where $\bt$ is a root in the Higgs bidoublet.  \label{condition2}

\item The states in $\bdm{(2,2)_\pm}$ and $\bdm{(1,1)_\pm}$ satisfy 
the zero-mode condition~(\ref{zm_cond:fermion}).  \label{condition3}
\end{enumerate}
We will search for $\cR$ that satisfies these conditions 
from the list in Appendix~\ref{decompList}. 
We focus on the cases of $G={\rm SO(5)},{\rm SU(4)},{\rm SO(7)}$(I), 
which have the $\cP_{\rm LR}$ symmetry. 

\begin{description}
\item[SO(5)] \mbox{}\\
There is no irreducible representation that satisfies 
the condition~\ref{condition1} among the list in Appendix~\ref{decompList:SO5}.   

\ignore{
We have shown in Sec.~\ref{zm:SO5} 
that only the case of $N=2$ has a bidoublet Higgs as a zero-mode. 
Since $(p_1,p_2)=(\pi,\pi)$ and the simple roots are 
$\alp_1=\bdm{e}^1-\bdm{e}^2$ and $\alp_2=\bdm{e}^2$, 
the zero-mode condition~(\ref{zm_cond:fermion}) becomes 
\be
 \frac{\pi(1-\chi_4\chi_6)}{2}-\pi k_2 = 0. \;\;\;\;\; (\mbox{mod $2\pi$}) 
\ee
As the simplest example, let us consider a case that 
the quarks are embedded into $\bdm{5_{2/3}}$ and $\bdm{5_{-1/3}}$. 
Here $\bdm{5}$ is decomposed as (\ref{decomp:SO5:5}), 
and $k_2=0,2$ for $\bdm{(2,2)}$ and $k_2=1$ for $\bdm{(1,1)}$. 
Thus $\bdm{5_{2/3}}$ and $\bdm{5_{-1/3}}$ should have the 6D chirality 
as $\chi_6=-1$ so that $\bdm{(2,2)}$ with $\chi_4=-1$ and 
$\bdm{(1,1)}$ with $\chi_4=+1$ have zero-modes. 
}

\item[SU(4)] \mbox{}\\
Only $\bdm{20'}$ satisfies the condition~\ref{condition1} 
among the list in Appendix~\ref{decompList:SU4}. 
The weights of $\bdm{20'}$ that form $\bdm{(2,2)}$ and $\bdm{(1,1)}$ are 
\bea
 \bdm{(2,2)_{\pm 2}} &:& 
 \brkt{\begin{array}{ccc} \frac{\bdm{e}^1-\bdm{e}^2}{\sqrt{2}}\pm\bdm{e}^3 & 
 \stackrel{\alp_R}{\longrightarrow} & \frac{\bdm{e}^1+\bdm{e}^2}{\sqrt{2}}\pm\bdm{e}^3 \\
 \uparrow_{\alp_L} & & \uparrow_{\alp_L} \\
 \frac{-\bdm{e}^1-\bdm{e}^2}{\sqrt{2}}\pm\bdm{e}^3 & 
 \stackrel{\alp_R}{\longrightarrow} & \frac{-\bdm{e}^1+\bdm{e}^2}{\sqrt{2}}\pm\bdm{e}^3
 \end{array}}, \nonumber\\
 \bdm{(1,1)_{\pm 4}} &:& \pm 2\bdm{e}^3, \;\;\;\;\;\;\;\;
 \bdm{(1,1)_0} \;\; : \;\; \bdm{0}. 
\eea
where the double signs correspond. 
Notice that the weights that form bidoublets are the same as 
the roots that form the Higgs bidoublets. 


When the Higgs bidoublet~$\bdm{(2,2)_{\pm 2}}$ appears as a zero-mode,  
one example of $(\bt,\mu_L,\nu_L)$ is chosen as 
\be
 (\bt,\mu_L,\nu_L) = 
 \brkt{\frac{\bdm{e}^1-\bdm{e}^2}{\sqrt{2}}\pm\bdm{e}^3,
 \frac{-\bdm{e}^1+\bdm{e}^2}{\sqrt{2}}\pm\bdm{e}^3,
 \frac{\bdm{e}^1-\bdm{e}^2}{\sqrt{2}}\mp\bdm{e}^3},  \label{choice:weight}
\ee
where the double signs correspond. 
Then $\{\mu_L-\bt,\mu_L,\mu_L+\bt\}$ and $\{\nu_L+\bt,\nu_L,\nu_L+\bt\}$ are the weights, 
but $\mu_L\pm 2\bt$ and $\nu_L\pm 2\bt$ are not. 
Therefore the condition~\ref{condition2} is satisfied (see (\ref{master_formula})). 

Since $(p_1,p_2,p_3)=(0,0,2n_P\pi/3)$ and the simple roots are 
$(\alp_1,\alp_2,\alp_3)=(\sqrt{2}\bdm{e}^1,-\frac{\bdm{e}^1}{\sqrt{2}}
-\frac{\bdm{e}^2}{\sqrt{2}}+\bdm{e}^3,\sqrt{2}\bdm{e}^2)$, 
the zero-mode condition~(\ref{zm_cond:fermion}) becomes 
\be
 \frac{\pi(2m_\omg+1-\chi_4\chi_6)}{N}-\frac{2n_Pk_2\pi}{N} = 0, 
 \;\;\;\;\;(\mbox{mod $2\pi$})
\ee
where $m_\omg=0,1,\cdots,N-1$. 
\ignore{
The decomposition of $\bdm{15}$ is given by (\ref{adj_decomp:SU4}), 
and 
\bea
 k_2=0 &:& \bdm{(2,2)_{+2}}, \nonumber\\
 k_2=1 &:& \bdm{(3,1)_0}, \;\; \bdm{(1,3)_0}, \;\; \bdm{(1,1)_0}, \nonumber\\
 k_2=2 &:& \bdm{(2,2)_{-2}}, 
\eea 
}
The decomposition of $\bdm{20'}$ is given by (\ref{decomp:SU4:20p}), and 
\bea
 k_2=0 &:& \bdm{(1,1)_{+4}}, \nonumber\\
 k_2=1 &:& \bdm{(2,2)_{+2}}, \nonumber\\
 k_2=2 &:& \bdm{(3,3)_0}, \;\; \bdm{(1,1)_0}, \nonumber\\
 k_2=3 &:& \bdm{(2,2)_{-2}}, \nonumber\\
 k_2=4 &:& \bdm{(1,1)_{-4}}. 
\eea
Thus the condition~\ref{condition3} is satisfied 
only when the model is compactified on $T^2/Z_3$. 
In fact, when $(N,n_P,m_\omg)=(3,1,0)$, 
the fermionic zero-modes from each 6D Dirac fermion contain 
\bea
 &&\cQ_L^{(+)}\in\bdm{(2,2)_{+2}}, \;\;\;\;\;
%
%
%
 \lmd_R^{(+)}\in\bdm{(1,1)_{+4}}, \nonumber\\
 &&\cQ_L^{(-)}\in\bdm{(2,2)_{-2}}, \;\;\;\;\;
%
%
%
 \lmd_R^{(-)}\in\bdm{(1,1)_{-4}}, 
\eea
and when $(N,n_P,m_\omg)=(3,2,2)$, they contain 
\bea
 &&\cQ_L^{(+)}\in\bdm{(2,2)_{-2}}, \;\;\;\;\;
%
%
%
 \lmd_R^{(+)}\in\bdm{(1,1)_{-4}}, \nonumber\\
 &&\cQ_L^{(-)}\in\bdm{(2,2)_{+2}}, \;\;\;\;\;
%
%
%
 \lmd_R^{(-)}\in\bdm{(1,1)_{+4}}. 
\eea
By introducing 4D localized fermions with appropriate quantum numbers 
to decouple unwanted zero-modes, 
the desired Yukawa couplings~(\ref{cL_yukawa2}) are obtained. 
For the other choices of $(N,n_P,m_\omg)$, we cannot obtain 
the necessary multiplets.

\item[SO(7) (I)] \mbox{}\\
The irreducible representations 
that satisfy the condition~\ref{condition1} 
among the list in Appendix~\ref{decompList:SO7} 
are $\bdm{21}$ and $\bdm{27}$. 
These also satisfy the condition~\ref{condition2}, 
but they cannot satisfy the condition~\ref{condition3} 
for any choice of $(N,n_P,m_\omg)$. 

\end{description}

\section{Higgs potential} \label{HiggsPotential}
In contrast to the 5D gauge-Higgs unification model, 
we have quartic couplings of the Higgs fields at tree level. 
The relevant terms in the 6D Lagrangian are
\bea
 \cL \eql -\frac{1}{4g_A^2}\tr\brkt{F^{(A)MN}F^{(A)}_{MN}}+\cdots \nonumber\\
 \eql -\frac{1}{2g_A^2(\pi R_1)^2}\tr\brkt{(\der^\mu A_z)^\dagger\der_\mu A_z}
 -\frac{1}{8g_A^2(\pi R_1)^4}\tr\brkt{\sbk{A_z,A_{\bar{z}}}^2}+\cdots. 
 \label{Higgs:relevantterms}
\eea
In this section, we calculate the classical Higgs potential~$V_{\rm tree}$ 
focusing on the Higgs bidoublets, 
which are relevant to the electroweak symmetry breaking. 
In the previous section, we have shown that 
only a model of $G=$SU(4) compactified on $T^2/Z_3$ has required zero-mode 
spectrum for the quarks. 
For the sake of completeness, however, 
we will also calculate $V_{\rm tree}$ in the other cases 
that have Higgs bidoublets. 
We have one Higgs bidoublet in the cases of 
SO(5) on $T^2/Z_2$, SU(4) on $T^2/Z_N$ ($N=3,4,6$), 
SO(7) (I) on $T^2/Z_N$ ($N=4,6$), 
and Sp(6) (II) or (III) on $T^2/Z_N$ ($N=3,4,6$), 
and we have two Higgs bidoublets in the case of SU(4) on $T^2/Z_2$.

\subsection{SO(5) case}
First we consider the SO(5) case. 
In this case, the roots that form the bidoublet are 
\be
 \brkt{\begin{array}{ccc} \bdm{e}^2 & \stackrel{\alp_R}{\longrightarrow} 
 & \bdm{e}^1 \\
 \uparrow_{\alp_L} && \uparrow_{\alp_L} \\
 -\bdm{e}^1 & \stackrel{\alp_R}{\longrightarrow} & -\bdm{e}^2 \end{array}}. 
 \label{bidoublet:SO5}
\ee
From (\ref{Higgs:relevantterms}), the kinetic terms of the zero-modes~$W_z^\bt$ 
in the 4D effective Lagrangian are 
\be
 \cL_{\rm eff} = -\frac{\cA}{2(g_A\pi R_1)^2}\sum_\bt(\der^\mu W_z^\bt)^*
 \der_\mu W_z^\bt+\cdots. 
\ee
We have used (\ref{normalize:generator}), and $\cA$ is the area of $T^2/Z_N$. 
Thus the canonically normalized Higgs bidoublet is defined as 
\be
 \cH = \begin{pmatrix} H_2^{2*} & H_1^1 \\ -H_2^{1*} & H_1^2 \end{pmatrix}
 \equiv \frac{\sqrt{\cA}}{\sqrt{2}g_A\pi R_1}
 \begin{pmatrix} W_z^{e^2} & W_z^{e^1} \\ -W_z^{-e^1} & W_z^{-e^2} 
 \end{pmatrix}. 
\ee
Then it follows that 
\bea
 A_z \eql \frac{\sqrt{2}g_A\pi R_1}{\sqrt{\cA}}\brkt{
 H_1^1 E_{e^1}+H_2^{2*} E_{e^2}+H_1^2 E_{-e^2}+H_2^{1*} E_{-e^1}},  \nonumber\\
 \sbk{A_z,A_{\bar{z}}} \eql \frac{2(g_A\pi R_1)^2}{\cA}\left[
 \brkt{\abs{H_1^1}^2-\abs{H_2^1}^2}H_1
 +\brkt{\abs{H_2^2}^2-\abs{H_1^2}^2}H_2 \right. \nonumber\\
 &&\hspace{20mm}
 +\left\{N_{e^1,e^2}\brkt{H_1^1 H_1^{2*}-H_2^1 H_2^{2*}}E_{\alp_L} \right. \nonumber\\
 &&\hspace{25mm}\left.\left.
 +N_{e^1,-e^2}\brkt{H_1^1 H_2^2-H_1^2 H_2^1}E_{\alp_R}+\hc\right\}\right], 
 \label{com:AzAbz}
\eea
where we have used (\ref{rel:Ns}). 
Hence, from (\ref{Higgs:relevantterms}), $V_{\rm tree}$ is calculated as
\bea
 V_{\rm tree} \eql \frac{\cA}{8g_A^2(\pi R_1)^4}\tr\brkt{\sbk{A_z,A_{\bar{z}}}^2} 
 \nonumber\\
 \eql \frac{g_A^2}{2\cA}\left[\brkt{\abs{H_1^1}^2-\abs{H_2^1}^2}^2
 +\brkt{\abs{H_2^2}^2-\abs{H_1^2}^2}^2 \right.\nonumber\\
 &&\hspace{10mm}\left.
 +2\abs{N_{e^1,e^2}}^2\abs{H_1^1H_1^{2*}-H_2^1H_2^{2*}}^2 
 +2\abs{N_{e^1,-e^2}}^2\abs{H_1^1H_2^2-H_1^2H_2^1}^2\right] \nonumber\\
 \eql \frac{g^2}{4}\brc{\brkt{H_2^\dagger H_2-H_1^\dagger H_1}^2
 +4\abs{\tl{H}_2^\dagger H_1}^2} 
 \nonumber\\
 \eql \frac{g^2}{4}\sbk{\brc{\tr\brkt{\cH^\dagger\cH}}^2-4\det\brkt{\cH^\dagger\cH}}, 
 \label{V_tree:rank2}
\eea
where $H_2\equiv (H_2^1,H_2^2)^t$ and $H_1\equiv (H_1^1,H_1^2)^t$ 
are the ${\rm SU(2)}_{\rm L}$ doublets with the hypercharge~$Y=1/2$. 
We have used that (\ref{def:4Dg}) with $\abs{\alp_L}^2=2$, 
and $\abs{N_{e^1,e^2}}^2=\abs{N_{e^1,-e^2}}^2=1$. 
The above result agrees with Eq.(7) in Ref.~\cite{Chang:2012iq}. 
The final expression in (\ref{V_tree:rank2}) is manifestly invariant 
under the transformation:~$\cH\to U_L\cH U_R^\dagger$ 
($U_L\in {\rm SU(2)}_{\rm L}$ and $U_R\in {\rm SU(2)}_{\rm R}$).

\subsection{Cases of rank-three groups} \label{Vtree:rank3}
Next we consider the cases of the rank-three groups. 
In these cases, the candidates for the zero-mode Higgs bidoublets consist of 
the following roots.  
\be
 \brkt{\begin{array}{ccc} \gm+\alp_L & \stackrel{\alp_R}{\longrightarrow} 
 & \gm+\alp_L+\alp_R \\
 \uparrow_{\alp_L} && \uparrow_{\alp_L} \\
 \gm & \stackrel{\alp_R}{\longrightarrow} & \gm+\alp_R \end{array}}, \;\;\;\;\;
 \brkt{\begin{array}{ccc} -\gm-\alp_R & \stackrel{\alp_R}{\longrightarrow} 
 & -\gm \\
 \uparrow_{\alp_L} && \uparrow_{\alp_L} \\
 -\gm-\alp_L-\alp_R & \stackrel{\alp_R}{\longrightarrow} & -\gm-\alp_L \end{array}}, 
 \label{bidoublet:rank3}
\ee
where $\gm=-\frac{\bdm{e}^1}{\sqrt{2}}-\frac{\bdm{e}^2}{\sqrt{2}}+\bdm{e}^3$ for SU(4), 
$\gm=-\bdm{e}^1+\bdm{e}^3$ for SO(7) (I), 
and $\gm=-\bdm{e}^2-\bdm{e}^3$ for Sp(6) (II) or (III). 
The canonically normalized Higgs bidoublets are defined as
\bea
 \cH_+ \eql \begin{pmatrix} H_{2+}^{2*} & H_{1+}^1 \\ -H_{2+}^{1*} & H_{1+}^2 
 \end{pmatrix} \equiv \frac{\sqrt{\cA}}{\sqrt{2}g_A\pi R_1}
 \begin{pmatrix} W_z^{\gm+\alp_L} & W_z^{\gm+\alp_L+\alp_R} \\
 -W_z^\gm & W_z^{\gm+\alp_R} \end{pmatrix}, \nonumber\\
 \cH_- \eql \begin{pmatrix} H_{2-}^{2*} & H_{1-}^1 \\ -H_{2-}^{1*} & H_{1-}^2 
 \end{pmatrix} \equiv \frac{\sqrt{\cA}}{\sqrt{2}g_A\pi R_1}
 \begin{pmatrix} W_z^{-\gm-\alp_R} & W_z^{-\gm} \\
 -W_z^{-\gm-\alp_L-\alp_R} & W_z^{-\gm-\alp_L} \end{pmatrix},  
\eea
where the signs in the suffixes denote the signs of the $U(1)_X$ charges. 
Then it follows that 
\bea
 A_z \eql \frac{\sqrt{2}g_A\pi R_1}{\sqrt{\cA}}
 \left(H_{1+}^1E_{\gm_{LR}}+H_{2+}^{2*}E_{\gm_L}+H_{1+}^2E_{\gm_R}+H_{2+}^{1*}E_\gm 
 \right. \nonumber\\
 &&\hspace{20mm}\left.
 +H_{1-}^1E_{-\gm}+H_{2-}^{2*}E_{-\gm_R}+H_{1-}^2E_{-\gm_L}+H_{2-}^{1*}E_{-\gm_{LR}}
 \right)+\cdots, \nonumber\\
 \sbk{A_z,A_{\bar{z}}} \eql \frac{2(g_A\pi R_1)^2}{\cA}
 \left[\brkt{\abs{H_{2+}^1}^2-\abs{H_{1-}^1}}\gm\cdot H
 +\brkt{\abs{H_{2+}^2}^2-\abs{H_{1-}^2}^2}\gm_L\cdot H 
 \right. \nonumber\\
 &&\hspace{20mm}
 +\brkt{\abs{H_{1+}^2}^2-\abs{H_{2-}^2}^2}\gm_R\cdot H
 +\brkt{\abs{H_{1+}^1}^2-\abs{H_{2-}^1}^2}\gm_{LR}\cdot H \nonumber\\
 &&\hspace{20mm}
 +\left\{N_{\gm_{LR},-\gm_R}
 \brkt{-H_{1+}^1H_{1+}^{2*}+H_{2-}^1H_{2-}^{2*}}E_{\alp_L} \right. \nonumber\\
 &&\hspace{25mm}
 +N_{\gm_{LR},-\gm_L}
 \brkt{-H_{1+}^1H_{2+}^2+H_{1-}^2H_{2-}^1}E_{\alp_R} \nonumber\\
 &&\hspace{25mm}
 +N_{\gm_L,-\gm}\brkt{-H_{2+}^1H_{2+}^{2*}+H_{1-}^1H_{1-}^{2*}}E_{\alp_L} 
 \nonumber\\
 &&\hspace{25mm}\left.\left. 
 +N_{\gm_R,-\gm}\brkt{-H_{1+}^2H_{2+}^1+H_{1-}^1H_{2-}^2}E_{\alp_R}
 +\hc\right\}\right]+\cdots, 
\eea
where $\gm_L\equiv\gm+\alp_L$, $\gm_R\equiv\gm+\alp_R$ 
and $\gm_{LR}\equiv\gm+\alp_L+\alp_R$, and the ellipses denote 
fields belonging to other multiplets, if any.  
\ignore{
Hence $V_{\rm tree}$ is calculated as 
\bea
 V_{\rm tree} \eql \frac{g_A^2}{2\cA}\left[
 \brkt{\abs{H_{2+}^2}^2-\abs{H_{1-}^1}^2}^2
 +\brkt{\abs{H_{2+}^1}^2-\abs{H_{1-}^2}^2}^2 \right. \nonumber\\
 &&\hspace{10mm}
 +\brkt{\abs{H_{1+}^2}^2-\abs{H_{2-}^1}^2}^2
 +\brkt{\abs{H_{1+}^1}^2-\abs{H_{2-}^2}^2}^2 \nonumber\\
 &&\hspace{10mm}
 +\brkt{\abs{H_{2+}^2}^2-\abs{H_{1-}^1}^2}
 \brkt{\abs{H_{2+}^1}^2-\abs{H_{1-}^2}^2} \nonumber\\
 &&\hspace{10mm}
 +\brkt{\abs{H_{2+}^2}^2-\abs{H_{1-}^1}^2}
 \brkt{\abs{H_{1+}^2}^2-\abs{H_{2-}^1}^2} \nonumber\\
 &&\hspace{10mm}
 +\brkt{\abs{H_{2+}^1}^2-\abs{H_{1-}^2}^2}
 \brkt{\abs{H_{1+}^1}^2-\abs{H_{2-}^2}^2} \nonumber\\
 &&\hspace{10mm}
 +\brkt{\abs{H_{1+}^2}^2-\abs{H_{1-}^1}^2}
 \brkt{\abs{H_{1+}^1}^2-\abs{H_{2-}^2}^2} \nonumber\\
 &&\hspace{10mm}
 +\abs{H_{1+}^1H_{1+}^{2*}+H_{2-}^1H_{2-}^{2*}
 -e^{i\zeta}\brkt{H_{2+}^1H_{2+}^{2*}+H_{1-}^1H_{1-}^{2*}}}^2 \nonumber\\
 &&\hspace{10mm}
 +\abs{H_{1+}^1H_{2+}^{1*}+H_{1-}^2H_{2-}^{2*}
 -e^{i\eta}\brkt{H_{1+}^2H_{2+}^{2*}+H_{1-}^1H_{2-}^{1*}}}^2, 
\eea
where $e^{i\zeta}\equiv N_{\gm_L,-\gm}/N_{\gm_{LR},-\gm_R}$ 
and $e^{i\eta}\equiv N_{\gm_R,-\gm}/N_{\gm_{LR},-\gm_L}$. 
}
After some calculations, we obtain 
\bea
 V_{\rm tree} \eql \frac{g^2}{2}\left[
 \brkt{\abs{H_{1+}}^2-\abs{H_{2-}}^2}^2
 +\brkt{\abs{H_{2+}}^2-\abs{H_{1-}}^2}^2 \right. \nonumber\\
 &&\hspace{5mm}
 +\abs{H_{1+}^\dagger \tl{H}_{2+}}^2+\abs{H_{1+}^\dagger \tl{H}_{2-}}^2
 +\abs{H_{1-}^\dagger \tl{H}_{2+}}^2+\abs{H_{1-}^\dagger \tl{H}_{2-}}^2 \nonumber\\
 &&\hspace{5mm}\left.
 -\abs{\tl{H}_{2+}^tH_{2-}}^2-\abs{\tl{H}_{1+}^tH_{1-}}^2
 +\abs{\tl{H}_{2+}^\dagger H_{1+}+\tl{H}_{2-}^\dagger H_{1-}}^2\right] \nonumber\\
 \eql \frac{g^2}{2}\left[
 \brc{\tr\brkt{\cH_+^\dagger\cH_+}}^2+\brc{\tr\brkt{\cH_-^\dagger\cH_-}}^2
 -\tr\brkt{\tl{\cH}_+^\dagger\tl{\cH}_+\cH_-^\dagger\cH_-} \right. \nonumber\\
 &&\hspace{5mm}\left. 
 -\tr\brkt{\cH_-^\dagger\tl{\cH}_+\tl{\cH}_+^\dagger\cH_-}
 -2\det\brkt{\cH_+^\dagger\cH_+}-2\det\brkt{\cH_-^\dagger\cH_-}\right]+\cdots, 
 \label{V_tree:rank3}
\eea
where $\tl{H}_{1,2+}^i\equiv\ep_{ij}H_{1,2+}^{j*}$, and 
$\tl{\cH}_\pm \equiv \sgm_2\cH_\pm^*\sgm_2$. 
We have used that 
\bea
 \gm\cdot\gm_{LR} \eql \gm_L\cdot\gm_R = 0, \nonumber\\
 \abs{\gm_{LR}}^2 \eql \abs{\gm_L}^2 = \abs{\gm_R}^2 = \abs{\gm}^2 = 2, \nonumber\\
 \gm\cdot\gm_L \eql \gm\cdot\gm_R 
 = \gm_L\cdot\gm_{LR} = \gm_R\cdot\gm_{LR} = \frac{\abs{\gm}^2}{2} = 1, \nonumber\\
 \abs{N_{\gm_{LR},-\gm_L}}^2 \eql \abs{N_{\gm_{LR},-\gm_R}}^2
 = \abs{N_{\gm_L,-\gm}}^2 = \abs{N_{\gm_R,-\gm}}^2 = \frac{\abs{\gm}^2}{2} = 1, 
 \nonumber\\
 \frac{N_{\gm_L,-\gm}}{N_{\gm_{LR},-\gm_R}} 
 \eql \frac{N_{\gm,\alp_L}^*}{N_{\gm_R,\alp_L}^*} 
 = \frac{N_{\gm,\alp_R}^*}{N_{\gm_L,\alp_R}^*}
 = \frac{N_{\gm_R,-\gm}}{N_{\gm_{LR},-\gm_L}}, 
\eea
which are followed by (\ref{rel:Ns}), (\ref{master_formula}) 
and the fact that $\alp_L\cdot\alp_R=0$ and $\sbk{E_{\alp_L},E_{\alp_R}}=0$. 
We have also chosen the phases of the Higgs fields so that
$N_{\gm_L,-\gm}/N_{\gm_{LR},-\gm_R}=-1$. 

The final expression in (\ref{V_tree:rank3}) is manifestly invariant 
under the transformation:~$\cH_\pm\to U_L\cH_\pm U_R$ 
($U_L\in {\rm SU(2)}_{\rm L}$ and $U_R\in {\rm SU(2)}_{\rm R}$). 
Except for the case of SU(4) on $T^2/Z_2$, one of the bidoublets~$\cH_\pm$ 
is absent due to the orbifold boundary conditions. 
In such cases, the model becomes a two-Higgs-doublet model. 
In contrast to the SO(5) case, 
the potential~(\ref{V_tree:rank3}) with $\cH_+=0$ or $\cH_-=0$ 
does not agree with (7) of Ref.~\cite{Chang:2012iq}.  
This is because they have assumed $\gm+\alp_L+\alp_R=-\gm$, 
which only holds in the SO(5) case. 


Finally we comment on the Higgs mass. 
We consider a case of SU(4) on $T^2/Z_3$. 
The tree-level Higgs potential~(\ref{V_tree:rank3}) becomes 
\bea
 V_{\rm tree} \eql \frac{g^2}{2}
 \sbk{\brc{\tr\brkt{\cH^\dagger\cH}}^2-2\det\brkt{\cH^\dagger\cH}} 
 \nonumber\\
 \eql \frac{g^2}{2}\brc{\brkt{H_1^\dagger H_1}^2+\brkt{H_2^\dagger H_2}^2
 +2\abs{\tl{H}_2^\dagger H_1}^2}, \label{V_tree}
\eea
where $\cH=(\tl{H}_2,H_1)$ is one of $\cH_\pm$. 
Since only the ${\rm U(1)}_{\rm em}$ neutral 
components~$H_1^2$ and $H_2^2$ can have nonzero VEVs, 
we focus on them. 
As discussed in Sec.~\ref{embed_quark}, 
we expect that $h_+\equiv(H_1^2+H_2^2)/\sqrt{2}$ has a tachyonic mass 
while $h_-\equiv(H_1^2-H_2^2)/\sqrt{2}$ does not at one-loop level. 
Including such mass terms, the potential becomes 
\be
 V = -m_+^2\abs{h_+}^2+m_-^2\abs{h_-}^2
 +\frac{g^2}{4}\brc{\brkt{\abs{h_+}^2+\abs{h_-}^2}^2
 +\brkt{h_+^*h_-+h_+h_-^*}^2}
 +\cdots,  
\ee
where $m_\pm>0$, 
and the ellipsis denotes terms involving the charged components. 
We can always redefine the phase of fields so that $\vev{h_+}>0$. 
Then, from the minimization condition for the potential, we obtain 
\be
 \vev{h_+} = \frac{\sqrt{2}m_+}{g}, \;\;\;\;\;
 \vev{h_-} = 0. 
\ee
Therefore, the alignment~(\ref{VEV:cH}) is actually achieved. 
Note that the one-loop induced quadratic terms do not have 
the ${\rm SU(2)}_{\rm L}\times{\rm SU(2)}_{\rm R}$ symmetry. 
Thus the custodial symmetry is broken in the Higgs sector. 
This does not cause a problem for the protection of the $\rho$ parameter 
at tree level as long as the relevant terms to the W and Z boson masses 
in (\ref{rel:WZmass}) have the custodial symmetry.  

The mass of the lightest neutral Higgs boson is 
\be
 m_H = g\abs{\vev{h_+}} = \frac{gv}{\sqrt{2}} = \sqrt{2}m_W, 
\ee
where $v$ is defined as $\vev{H_1^2}=\vev{H_2^2}=v/2>0$, 
and we have used (\ref{expr:m_W}) at the last equality. 
We expect that the deviation from the observed value~$m_H\simeq 125$~GeV 
is explained by quantum corrections~\footnote{
Although the quantum correction to the Higgs quartic couplings is divergent, 
the ratio~$m_H/m_W$ is proven to be finite and calculable in Ref.~\cite{Lim:2014tua}.  
}. 

\section{Summary} \label{summary}
We have investigated 6D gauge-Higgs unification models 
compactified on $T^2/Z_N$ ($N=2,3,4,6$) that have the custodial symmetry. 
The gauge group is assumed to be ${\rm SU(3)}_C\times G\times {\rm U(1)}_Z$, 
where $G$ is a simple group. 
Since $G$ includes ${\rm SU(2)}_{\rm L}\times{\rm SU(2)}_{\rm R}$, 
its rank must be more than one. 
The Higgs fields originate from the extra-dimensional components of 
the $G$ gauge field. 
In contrast to 5D models~\cite{Agashe:2004rs,Hosotani:2006qp,Hosotani:2008tx}, 
we have at least two Higgs doublets. 
Thus their VEVs need to be aligned as (\ref{VEV:cH}) 
to preserve the custodial symmetry. 
This severely constrains the structure of models. 

In order to select candidates for viable models, 
we demanded the following requirements. 
\begin{itemize}
\item The model has a scalar bidoublet zero-mode as the Higgs fields. 

\item The bosonic sector has a symmetry under a parity~$\cP_{\rm LR}$ 
that exchanges SU(2)${}_{\rm L}$ and SU(2)${}_{\rm R}$ 
in order to protect the $Zb_L\bar{b}_L$ coupling 
against a large deviation induced by mixing with the KK modes. 

\item The quark fields are embedded into 6D fermions  
in such a way that they couple to the Higgs bidoublet~$\cH$ only through 
a combination~$\cH+\sgm_2\cH^*\sgm_2$. 

\item The representation~$\cR$ that the 6D fermions belong to 
provides a right size group factor to realize the top Yukawa coupling constant. 
\end{itemize}
The third requirement is demanded in order for the Higgs VEVs 
to be aligned as (\ref{VEV:cH}). 
The third and the fourth requirements can be achieved 
if $\cR$ satisfies the three conditions in Sec.~\ref{avl_rep}. 
Our results are summarized in Table~\ref{summary_table}. 
In the cases with blank, 
there is no choice of the orbifold boundary conditions 
so that $G$ is broken to 
${\rm SU(2)}_{\rm L}\times{\rm SU(2)}_{\rm R}(\times {\rm U(1)}_X)$. 
There is only one candidate that satisfies the above requirements 
if we restrict ourselves to the cases that ${\rm rank}\,G\leq 3$ and $\dim\cR<30$.  
It is the case of $G=$SU(4), $N=3$ and $\cR=\bdm{20'}$. 
Namely, the model is 6D ${\rm SU(3)}_C\times{\rm U(4)}$ gauge theory 
compactified on $T^2/Z_3$, 
and the top and bottom quarks are embedded into 
the symmetric traceless rank-2 tensor of SO(6).\footnote{
Since the zero-mode spectrum is non-chiral in this case, 
the chiral structure of the effective theory must be attributed to  
the chiral field content of the localized fermions at the fixed point. 
} 
\begin{table}[t, b]
\begin{center}
\begin{tabular}{|c||c|c|c|c|c|c|c|}
 \hline \rule[-2mm]{0mm}{7mm}  & SO(5) & G${}_2$ & 
 SU(4) & \multicolumn{2}{|c|}{SO(7)} & \multicolumn{2}{|c|}{Sp(6)} \\ \cline{5-8}
 & & & & (I) & (II), (III) & (I) & (II), (III) \\ \hline
 $T^2/Z_2$ & 1 (S) & 0 & 2 (S) &  &  &  &  \\ \hline 
 $T^2/Z_3$ &  &  & 1 (S) $\checkmark$ &  & 0 &  & 1 \\ \hline
 $T^2/Z_4$ & 0 (S) & 0 & 1 (S) & 1 (S) & 0 & 0 (S) & 1 \\ \hline 
 $T^2/Z_6$ & 0 (S) & 0 & 1 (S) & 1 (S) & 0 & 0 (S) & 1 \\ \hline
 \end{tabular}
\end{center}
\caption{Summary of the results. 
The numbers denote those of the Higgs bidoublets. 
(I), (II) and (III) represent 
three different ways of choosing 
the ${\rm SU(2)}_{\rm L}\times{\rm SU(2)}_{\rm R}$ subgroup in Sec.~\ref{ZM:rank3}. 
"(S)" indicates that the spectrum is symmetric  
under SU(2)${}_{\rm L}$ $\leftrightarrow$ SU(2)${}_{\rm R}$.   
The check mark is added to a case that there is an appropriate embedding of quarks 
into 6D fermions. 
}  
\label{summary_table}
\end{table}
We have focused on the third generation quarks to restrict $G$, $N$ and $\cR$. 
Embeddings of other fermions are much less constrained. 

There are many issues that we have not discussed in this paper. 
We have approximated the mode functions of the W and Z bosons 
as constants. 
However, after the electroweak symmetry is broken, 
they are no longer constants and depend on $z$. 
This $z$-dependence causes the deviation of the $\rho$ parameter 
and the $Zb_L\bar{b}_L$ coupling from the standard model values. 
We have to check that the custodial symmetry actually suppresses these deviations 
by using the exact mode functions. 
We should also calculate the one-loop effective potential 
to check that the vacuum alignment~(\ref{VEV:cH}) is actually achieved, 
and to evaluate the Higgs mass spectrum. 
The moduli stabilization in the gauge-Higgs unification is 
also an important subject~\cite{Sakamura:2010ju,Maru:2010ap}. 
It is interesting to investigate this subject 
in the presence of a background magnetic flux on $T^2/Z_N$. 
All these issues are left for our future works.

\subsection*{Acknowledgements}
This work was supported in part by 
Grant-in-Aid for Scientific Research (C) No.25400283  
from Japan Society for the Promotion of Science (Y.S.).

\appendix

\section{Cartan-Weyl basis} \label{CartanWeyl}
The generators of a simple group~$G$ whose rank is $r$ in the Cartan-Weyl basis are
$H_i$ ($i=1,\cdots,r$) and $E_\alp$, which satisfy 
\bea
 H_i^\dagger \eql H_i, \;\;\;\;\;
 E_\alp^\dagger = E_{-\alp}, \nonumber\\
 \sbk{H_i,H_j} \eql 0, \;\;\;\;\;
 \sbk{H_i,E_\alp} = \alp_i E_\alp, \nonumber\\
 \sbk{E_\alp,E_\bt} \eql N_{\alp,\bt}E_{\alp+\bt}, 
 \;\;\;\;\;
 \sbk{E_\alp,E_{-\alp}} = \alp\cdot H, 
\eea
where $\alp,\bt$ are the root vectors, and $\alp\neq -\bt$. 
A complex constant $N_{\alp,\bt}$ is nonzero only when $\alp+\bt$ is a root, 
and satisfies the following equations. 
\be
 N_{\alp,\bt} = -N_{\bt,\alp} = -N^*_{-\alp,-\bt} 
 = N_{\bt,-\alp-\bt} = N_{-\alp-\bt,\alp}.  \label{rel:Ns}
\ee
For a series of the weights~$\{\mu-q\alp,\cdots,\mu-\alp,\mu,\mu+\alp,\cdots,\mu+p\alp\}$, 
where $p$ and $q$ are integers 
and neither $\mu-(q+1)\alp$ nor $\mu+(p+1)\alp$ is a weight, 
it follows that
\be
 \frac{2\alp\cdot\mu}{\abs{\alp}^2} = q-p, \;\;\;\;\;
 \abs{N_{\alp,\mu}}^2 = \frac{p(q+1)\abs{\alp}^2}{2},  \label{master_formula}
\ee
where a complex constant~$N_{\alp,\mu}$ is defined as
$E_\alp|\mu\rangle=N_{\alp,\mu}|\mu+\alp\rangle$. 
The generators are normalized as 
\be
 \tr(H_iH_j) = \dlt_{ij}, \;\;\;\;\;
 \tr(H_iE_\alp) = 0, \;\;\;\;\;
 \tr(E_\alp E_\bt) = \dlt_{\alp,-\bt}.  \label{normalize:generator}
\ee

\section{Orbifold boundary conditions} \label{general_obBC}
The orbifold~$T^2/Z_N$ is defined by identifying points of $\mathbb{R}^2$ 
by a discrete group~$\Gm$ which is generated by three descrete 
transformations~$\cO_1$:~$z\to z+1$, 
$\cO_\tau$:~$z\to z+\tau$ and $\cO_\omg$:~$z\to \omg z$. 
Field values of a 6D field at $\Gm$-equivalent points 
must be related to each other through gauge transformations~\footnote{
More properly, they are related through automorphisms 
of the Lie algebra of $G$. 
For simplicity, we do not consider a case of outer automorphisms~\cite{Hebecker:2001jb}. 
} 
in order for the Lagrangian to be single-valued on $T^2/Z_N$. 
Thus the most general orbifold boundary conditions are given by 
\bea
 A_M(x,z+1) \eql T_1A_M(x,z)T_1^{-1}, \;\;\;\;\;
 B_M^Z(x,z+1) = B_M^Z(x,z), \nonumber\\
 \Psi_{\chi_6}(x,z+1) \eql e^{i\vph_1}T_1\Psi_{\chi_6}(x,z), 
 \label{ob_BC:general1}
\eea
for the translation~$\cO_1$, 
\bea
 A_M(x,z+\tau) \eql T_\tau A_M(x,z)T_\tau^{-1}, \;\;\;\;\;
 B_M^Z(x,z+\tau) = B_M^Z(x,z), \nonumber\\
 \Psi_{\chi_6}(x,z+\tau) \eql e^{i\vph_\tau}T_\tau\Psi_{\chi_6}(x,z), 
 \label{ob_BC:general2}
\eea
for the translation~$\cO_\tau$, and 
\bea
 A_\mu(x,\omg z) \eql P A_\mu(x,z) P^{-1}, \;\;\;\;\;
 A_z(x,\omg z) = \omg^{-1}P A_z(x,z)P^{-1}, \nonumber\\
 B_\mu^Z(x,\omg z) \eql B_\mu^Z(x,z), \;\;\;\;\;
 B_z^Z(x,\omg z) = \omg^{-1}B_z^Z(x,z), \nonumber\\
 \Psi_{\chi_4,\chi_6}(x,\omg z) \eql 
 \omg^{-\frac{\chi_4\chi_6}{2}}e^{i\vph_\omg}P\Psi_{\chi_4,\chi_6}, 
 \label{ob_BC:general3}
\eea
for the $Z_N$ twist~$\cO_\omg$. 
Matrices~$T_1$, $T_\tau$ and $P$ are elements of $G$, 
and $\vph_1$ and $\vph_\tau$ are the Scherk-Schwarz phases. 
A factor~$\omg^{-1}$ and $\omg^{-\frac{\chi_4\chi_6}{2}}$ in (\ref{ob_BC:general3}) 
appears because $A_z$, $B_z^Z$ and $\Psi_{\chi_4,\chi_6}$ 
are charged under the rotation in the extra-dimensional space. 
Since $(\omg^{-\frac{\chi_4\chi_6}{2}})^N=-1$, 
the phase~$\vph_\omg$ is determined so that 
\be
 e^{iN\vph_\omg}P^N = -\id. \label{cond_for_vphomg}
\ee
The matrices~$T_1$, $T_\tau$ and $P$ satisfy the relations, 
\bea
 \sbk{T_1,T_\tau} \eql 0, \;\;\;\;\;
 P^N = \id, \nonumber\\
 P^{-1}T_1P \eql \begin{cases} T_1^{-1} & (N=2) \\
 T_\tau^{-1}T_1^{-1} & (N=3) \\
 T_\tau^{-1} & (N=4) \\ T_\tau^{-1}T_1 & (N=6) \end{cases}, \;\;\;\;\;
 P^{-1}T_\tau P = \begin{cases} T_\tau^{-1} & (N=2) \\
 T_1 & (N=3,4,6) \end{cases},  \label{rels:TP}
\eea
which reflect the properties of $\cO_1$, $\cO_\tau$ and $\cO_\omg$. 

Here we perform a gauge transformation, 
\be
 A_M \to UA_MU^{-1}+iU\der_M U^{-1}, \;\;\;\;\;
 \Psi \to U\Psi, 
\ee
where 
\be
 U(z) \equiv \exp\brc{-\frac{\Im(\tau\bar{z})}{\Im\tau}\ln T_1
 -\frac{\Im z}{\Im\tau}\ln T_\tau},  \label{def:U}
\ee
Using (\ref{rels:TP}), we can show that 
\bea
 U(z+1) \eql U(z)T_1^{-1}, \;\;\;\;\;
 U(z+\tau) = U(z)T_\tau^{-1},  \nonumber\\
 P^{-1}U(\omg z)P \eql U(z), \;\;\;\;\;
 P^{-1}\brkt{iU\der_z U^{-1}}P = \omg^{-1}\brkt{iU\der_z U^{-1}}. 
\eea
Thus, the matrices~$T_1$ and $T_\tau$ 
in (\ref{ob_BC:general1}) and (\ref{ob_BC:general2}) 
can be absorbed by this gauge transformation, 
while the conditions in (\ref{ob_BC:general3}) are unchanged. 
Since we need the fermionic zero-modes, 
we assume that $\vph_1=\vph_\tau=0$ for the fermion that the quarks are embedded. 
Then the orbifold boundary conditions are reexpressed as
(\ref{orbifold_BC:gauge}) and (\ref{ob_cond:fermion}).

\section{Decomposition of $\bdm{G}$ representations} \label{decompList}
Here we list various representations of $G=$SO(5),SU(4),SO(7), 
and their irreducible decompositions to multiplets of 
the ${\rm SU(2)}_{\rm L}\times {\rm SU(2)}_{\rm R}(\times {\rm U(1)}_X)$ subgroup. 

Each representation is specified by the Dynkin coefficients~$m_i$ ($i=1,\cdots,r$),  
and the highest weight is expressed as $\mu_{\rm max}=\sum_im_i\mu_i$, 
where $\mu_i$ are the fundamental weights.  
The dimension of the representation is calculated by the Weyl dimension formula: 
\be
 \dim\cR = \prod_l\frac{\sum_i(m_i+1)l_i\abs{\alp_i}^2}{\sum_il_i\abs{\alp_i}^2}, 
 \label{dim_formula}
\ee
where $\alp_i$ are the simple roots, and $l_i$ are numbers such that 
$\sum_il_i\alp_i$ are positive roots. 
We focus on irreducible representations whose dimensions are less than 30 
in the following.\footnote{ 
The irreducible decompositions of other representations 
and the weights of each representation are easily obtained 
by using LieART~\cite{Feger:2012bs}. } 

\subsection{SO(5)} \label{decompList:SO5}
The simple roots are $(\alp_1,\alp_2)=(\bdm{e}^1-\bdm{e}^2,\bdm{e}^2)$, 
and the fundamental weights are $(\mu_1,\mu_2)=(\bdm{e}^1,\frac{\bdm{e}^1+\bdm{e}^2}{2})$. 
The dimension formula~(\ref{dim_formula}) becomes 
\be
 \dim\cR = \frac{1}{6}(m_1+1)(m_2+1)(m_1+m_2+2)(2m_1+m_2+3). 
\ee
The decompositions to the irreducible representations of 
${\rm SU(2)}_{\rm L}\times {\rm SU(2)}_{\rm R}$ are as follows. 
\begin{description}
\item[$\bdm{[m_1,m_2]=[1,0]}$] 
\be
 \bdm{5 = (2,2)+(1,1)}.  \label{decomp:SO5:5}
\ee

\item[$\bdm{[m_1,m_2]=[0,1]}$] 
\be
 \bdm{4 = (2,1)+(1,2)}. 
\ee

\item[$\bdm{[m_1,m_2]=[2,0]}$] 
\be
 \bdm{14 = (3,3)+(2,2)+(1,1)}. 
\ee

\item[$\bdm{[m_1,m_2]=[1,1]}$] 
\be
 \bdm{16 = (3,2)+(2,3)+(2,1)+(1,2)}. 
\ee

\item[$\bdm{[m_1,m_2]=[0,2]}$] \mbox{}\\
This is the adjoint representation and decomposed as (\ref{IRdecomp:SO5}).

\item[$\bdm{[m_1,m_2]=[0,3]}$] 
\be
 \bdm{20 = (4,1)+(3,2)+(2,3)+(1,4)}. 
\ee
\end{description}

\subsection{SU(4)} \label{decompList:SU4}
The simple roots are $(\alp_1,\alp_2,\alp_3)=(\sqrt{2}\bdm{e}^1,
-\frac{\bdm{e}^1}{\sqrt{2}}-\frac{\bdm{e}^2}{\sqrt{2}}+\bdm{e}^3,\sqrt{2}\bdm{e}^2)$, 
and the fundamental weights are $(\mu_1,\mu_2,\mu_3)=
(\frac{\bdm{e}^1}{\sqrt{2}}+\frac{\bdm{e}^3}{2},\bdm{e}^3,
\frac{\bdm{e}^2}{\sqrt{2}}+\frac{\bdm{e}^3}{2})$. 
The dimension formula~(\ref{dim_formula}) becomes 
\bea
 \dim\cR \eql \frac{1}{12}(m_1+1)(m_2+1)(m_3+1)(m_1+m_2+2) \nonumber\\
 &&\times(m_2+m_3+2)(m_1+m_2+m_3+3). 
\eea
The decompositions to the irreducible representations of 
${\rm SU(2)}_{\rm L}\times {\rm SU(2)}_{\rm R}\times {\rm U(1)}_X$ are as follows. 
\begin{description}
\item[$\bdm{[m_1,m_2,m_3]=[1,0,0]}$] 
\be
 \bdm{4 = (2,1)_{+1}+(1,2)_{-1}}. 
\ee

\item[$\bdm{[m_1,m_2,m_3]=[0,1,0]}$] 
\be
 \bdm{6 = (2,2)_0+(1,1)_{+2}+(1,1)_{-2}}.  \label{decomp:SU4:6}
\ee

\item[$\bdm{[m_1,m_2,m_3]=[0,0,1]}$] 
\be
 \bdm{\bar{4} = (2,1)_{-1}+(1,2)_{+1}}. 
\ee

\item[$\bdm{[m_1,m_2,m_3]=[1,0,1]}$] \mbox{}\\
This is the adjoint representation and 
decomposed as (\ref{adj_decomp:SU4}). 

\item[$\bdm{[m_1,m_2,m_3]=[0,1,1]}$] 
\be
 \bdm{20 = (3,2)_{-1}+(2,3)_{+1}+(2,1)_{+1}+(2,1)_{-3}
 +(1,2)_{+3}+(1,2)_{-1}}. 
\ee

\item[$\bdm{[m_1,m_2,m_3]=[0,2,0]}$]
\be
 \bdm{20' = (3,3)_0+(2,2)_{+2}+(2,2)_{-2}+(1,1)_{+4}+(1,1)_{-4}+(1,1)_0}. 
 \label{decomp:SU4:20p}
\ee

\item[$\bdm{[m_1,m_2,m_3]=[1,1,0]}$]
\be
 \bdm{\overline{20} = (3,2)_{+1}+(2,3)_{-1}+(2,1)_{+3}+(2,1)_{-1}
 +(1,2)_{+1}+(1,2)_{-3}}. 
\ee
 
\item[$\bdm{[m_1,m_2,m_3]=[0,0,3]}$]
\be
 \bdm{20'' = (4,1)_{-3}+(3,2)_{-1}+(2,3)_{+1}+(1,4)_{+3}}. 
\ee

\item[$\bdm{[m_1,m_2,m_3]=[3,0,0]}$]
\be
 \bdm{\overline{20}'' = (4,1)_{+3}+(3,2)_{+1}+(2,3)_{-1}+(1,4)_{-3}}. 
\ee

\end{description}

\subsection{SO(7)} \label{decompList:SO7}
The simple roots are $(\alp_1,\alp_2,\alp_3)=(\bdm{e}^1-\bdm{e}^2,\bdm{e}^2-\bdm{e}^3,
\bdm{e}^3)$, and the fundamental weights are 
$(\mu_1,\mu_2,\mu_3)=(\bdm{e}^1,\bdm{e}^1+\bdm{e}^2,
\frac{\bdm{e}^1+\bdm{e}^2+\bdm{e}^3}{2})$. 
The dimension formula~(\ref{dim_formula}) becomes 
\bea
 \dim\cR \eql \frac{1}{720}(m_1+1)(m_2+1)(m_3+1)(m_1+m_2+2)(m_2+m_3+2)(2m_2+m_3+3) 
 \nonumber\\
 &&\times (m_1+m_2+m_3+3)(m_1+2m_2+m_3+4)(2m_1+2m_2+m_3+5). 
\eea
The ${\rm SU(2)}_{\rm L}\times {\rm SU(2)}_{\rm R}$ subgroup is chosen as 
$(\alp_L,\alp_R)=(\bdm{e}^1+\bdm{e}^2,\bdm{e}^1-\bdm{e}^2)$. 
The decompositions to the irreducible representations of 
${\rm SU(2)}_{\rm L}\times {\rm SU(2)}_{\rm R}\times {\rm U(1)}_X$ are as follows. 
\begin{description}
\item[$\bdm{[m_1,m_2,m_3]=[1,0,0]}$]
\be
 \bdm{7 = (2,2)_0+(1,1)_{+1}+(1,1)_{-1}+(1,1)_0}.  \label{decomp:SO7:7}
\ee

\item[$\bdm{[m_1,m_2,m_3]=[0,1,0]}$] \mbox{}\\
This is the adjoint representation and decomposed as (\ref{adj_decomp:SO7}). 

\item[$\bdm{[m_1,m_2,m_3]=[0,0,1]}$]
\be
 \bdm{8 = (2,1)_{+1/2}+(2,1)_{-1/2}+(1,2)_{+1/2}+(1,2)_{-1/2}}. 
\ee

\item[$\bdm{[m_1,m_2,m_3]=[2,0,0]}$]
\bea
 \bdm{27} \!\!\!&\bdm{=}\!\!\!& \bdm{(3,3)_0+(2,2)_{+1}+(2,2)_{-1}
 +(2,2)_0+(1,1)_{+2}} \nonumber\\
 &&\bdm{+(1,1)_{+1}+(1,1)_{0}+(1,1)_0+(1,1)_{-1}+(1,1)_{-2}}. 
\eea

\end{description}


\end{document}